\newcommand{\coloronline}[1]{{}} 
\documentclass[twocolumn,
    superscriptaddress,
    preprintnumbers,
    amsmath,
    amssymb,
    nofootinbib,
    tightenlines, 
    nobibnotes,pra,
    floatfix]{revtex4-1}
\usepackage{times}
\usepackage{mathrsfs}
\usepackage{braket}
\usepackage{graphicx}
\usepackage{bm}
\usepackage[colorlinks=true,linkcolor=blue]{hyperref}
\usepackage{natbib}
\DeclareMathOperator{\tr}{tr}
\newcommand{\av}[1]{\langle #1\rangle}

\usepackage{color}

\DeclareMathOperator\erf{erf}
\DeclareMathOperator\erfc{erfc}
\DeclareMathOperator\re{Re}
\DeclareMathOperator\im{Im}

\begin{document}

\title{Time-energy filtering of single electrons in ballistic waveguides}
\author{Elina Locane}
\affiliation{Dahlem Center for Complex Quantum Systems and Institut f\"ur Theoretische Physik, Freie Universit\"at Berlin, Arnimallee 14, 14195 Berlin, Germany}
\author{Piet W. Brouwer}
\affiliation{Dahlem Center for Complex Quantum Systems and Institut f\"ur Theoretische Physik, Freie Universit\"at Berlin, Arnimallee 14, 14195 Berlin, Germany}
\author{Vyacheslavs Kashcheyevs}
\email{Corresponding author: slava@latnet.lv}
\affiliation{Department of Physics, University of Latvia, Jelgavas street 3, LV-1004 Riga, Latvia}

\begin{abstract}
Characterizing distinct electron wave packets is a basic task for solid-state electron quantum optics with applications in quantum metrology and sensing. A important circuit element for this task is a non-stationary potential barrier that enables backscattering of chiral particles  depending on their energy and time of arrival. Here we solve the quantum mechanical problem of single-particle scattering by a ballistic constriction in an fully depleted quantum Hall system under spatially uniform but time-dependent electrostatic potential modulation.  The result describes electrons distributed in time-energy space according to a modified Wigner quasiprobability distribution and scattered with an energy-dependent transmission probability that characterizes constriction in the absence of modulation. Modification of the incoming Wigner distribution due to external time-dependent potential simplifies in case of linear time-dependence and admits semiclassical interpretation.
Our results support a recently proposed and implemented method for measuring time and energy distribution of solitary electrons 
as a quantum tomography technique, and offer new paths for experimental exploration of on-demand sources of coherent electrons.  
\end{abstract}

\maketitle
\section{Introduction}

Electron quantum optics is a relatively new field, aiming to reproduce quantum optics experiments with electron wave packets instead of photons \cite{degiovanni_rev_2011,
bocquillon_rev,
bauerle_2018,feve_bocq_rev}. It offers the prospects of probing the interaction between just a few electrons, as well as studying phenomena on the scale of electron coherence time. The field has potentially promising applications in signal processing \cite{feve_rev} and quantum sensing \cite{demkowicz_2012}.

One of the crucial ingredients that has made the investigation of single electron excitations in ballistic waveguides possible is the advent of devices that emit ordered streams of electrons with sufficient separation between individual particles~\cite{pothier_1992,marcus_1999,feve_2007,kaestner_2007,fujiwara_2008,pekola_2008,
kaestner_2008,blumenthal_2008,kaestner_2009,chorley_2012,npl_ghz_2013,dubois_2013,
ubbelohde_2015,dHollsoy_2015,Roussely2018}. Techniques to characterise quantum properties of electrical current have been adapted from photon quantum optics. Statistical properties of the source and the exchange statistics of the particles can be probed using intensity interferometers such as the Hanbury-Brown-Twiss and Hong-Ou-Mandel interferometers \cite{henny_1999,oliver_1999,kiesel_2002,bocquillon_2012,Bocquillon2013,Ferraro2018}, while coherence, entanglement, and the wave-like nature of particles can be probed using amplitude interferometers such as the Mach-Zehnder interferometer \cite{heiblum_2003,heiblum_2006,heiblum_2007}. 

\begin{figure}
\center
\includegraphics[width=0.95\columnwidth]{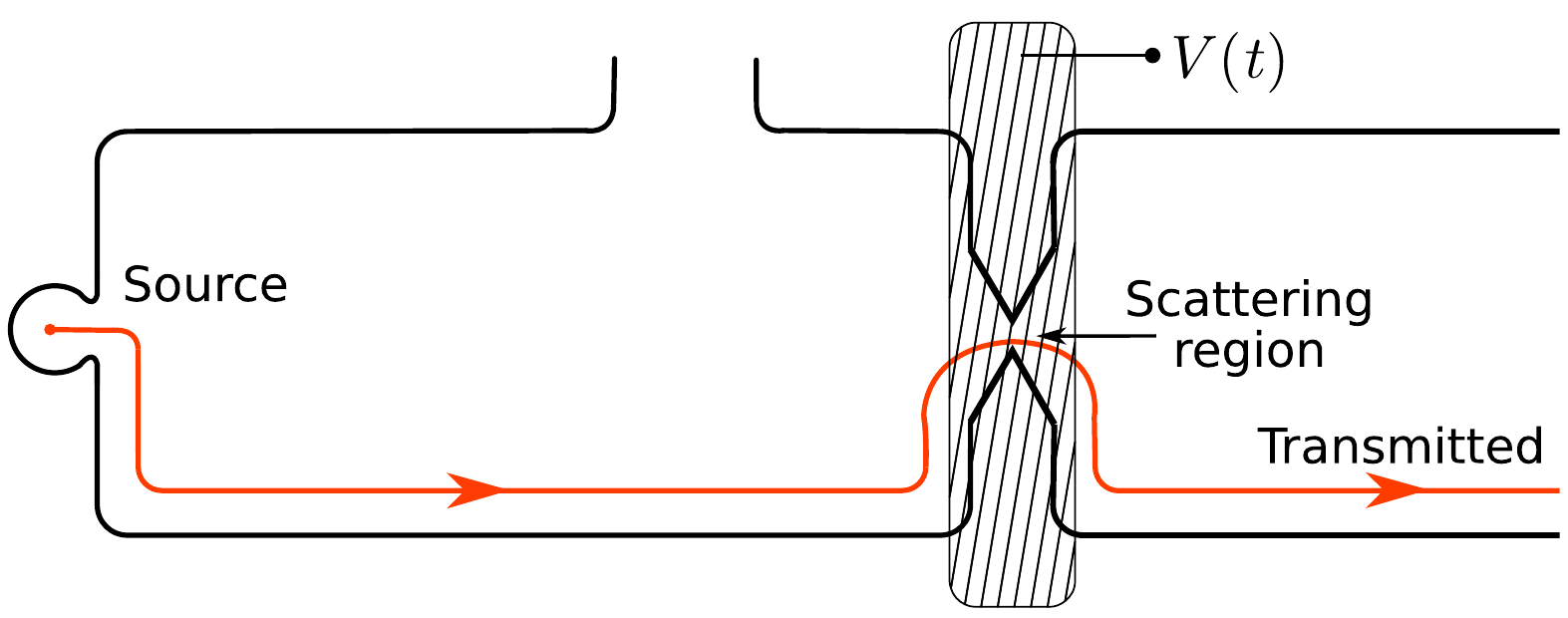} 
\caption{\coloronline{(color online)} A sketch of the setup: 
a chiral edge states (orange) in a quantum Hall system pass through  a constriction where  backscattering between the edges is possible. Individual electronic excitations are launched from the source (left) and get either transmitted or reflected.
The scattering region is electrostatically gated and the energy-dependent transmission probability $T(E)$ can be modulated by the voltage on the gate (shaded shape). It is assumed that the potential $V(t)$ created by the gate voltage modulation is spatially uniform in  the scattering region where backscattering between the edge channels takes place. Tuning the time- and energy- dependence of the transmission function $T(E-V(t))$ by appropriately designed time-dependence of the gate potential $V(t)$ enables tomography --- the measurement of joint energy-time content --- of the wavepacket.}
\label{fig:setup_exp}
\end{figure}

However, most of the existing electron quantum optics experiments have focused on single-electron sources that emit electrons close to the Fermi sea~\cite{Bocquillon2014,bauerle_2018}. These excitations can be accessed by perturbing the Fermi sea through the application of periodic gate potentials allowing for quantum state reconstruction through correlation measurements between the unknown and the reference signal \cite{Grenier2011,glattli_2014,Marguerite2017}. Such an approach is not possible in the case of ``high-energy'' electrons \cite{kaestner2010d,fletcher_detect_2013,ubbelohde_2015,waldie_detect_2015,kataoka_detect_2016,kataoka_pss} emitted far above (tens of meV)  the Fermi sea, because they do not overlap with perturbations around the Fermi energy. These excitations provide a new kind of quasi-particle whose coherence properties are largely unexplored.

A recent work by Fletcher {\it et al.} \cite{fletcher_2018} addresses the characterization of these ``high-energy'' electrons by using an energy barrier that is tuned to match the energy of the quasiparticles. The authors propose and implement an  electron tomography protocol that reconstructs the joint energy-time quasi-probability distribution $p(E,t)$ 
 of the incoming electrons by measuring the transmitted charge through an energy barrier whose height is varied linearly in time. 
The experimental setup can be described by a model shown schematically in Fig.\ \ref{fig:setup_exp}, where the time-dependent energy barrier is modelled with a static constriction subject to a (locally) uniform gate voltage $V(t)=V_0+\alpha \, t$.
The two parameters of the linear modulation (the offset in energy $V_0$ and the slope $\alpha$ in the time-energy plane) provide a two-dimensional map of the charge $Q(V_0,\alpha)$, that is used to infer $p(E,t)$ using inversion techniques from tomographic image processing~\cite{fletcher_2018}.   The key relation connecting the properties of the incoming wave-packets with the measured signal can be written as 
\begin{align}
 Q & = e \int dE \, dt\, T(E) \, p(E+V(t),t), 
  \label{eq:Q_semicl}
\end{align}
where $T(E)$ is the energy-dependent transmission probability of the constriction in the absence of the gate voltage modulation.
\footnote{Strictly speaking, in Eq.\ (\ref{eq:Q_semicl}) $V(t)$ should be replaced by the integral of the electric field for electrons traveling along the edge, which need not be equal to $V(t)$ for a time-dependent gate potential, see Sec.\ \ref{sec:derivation}.}

Equation \eqref{eq:Q_semicl} has been previously derived classically \cite{kataoka_pss} by considering scattering of a statistical ensemble of electrons with simultaneously well-defined energy and time of arrival, impinging on a scattering barrier with a transmission probability $T(E,t)$. Due to gauge invariance under uniform modulation, shift in voltage is equivalent to shift in energy, $T(E,t)=T(E-V(t),t)$, giving Eq.~\eqref{eq:Q_semicl}. In this classical picture $p(E,t)$ is simply the joint probability density characterizing the electrons emitted by the source.

In this work, we derive a general expression for the transmitted charge for an arbitrary time-dependence of $V(t)$ and show that the distribution $p(E,t)$ in a general case needs to be replaced by a suitably modified Wigner function. We also show that if the gate voltage $V(t)$ has a linear dependence on time, no modification of the Wigner function is necessary and  Eq.~\eqref{eq:Q_semicl} is valid if we take $p(E,t)$ to be the Wigner function of the incoming electron $\rho_{\text{in}}(E,t)$. This confirms that the protocol implemented in \cite{fletcher_2018} can be used not only for classical but also for quantum tomography, i.e., quantum state reconstruction. We argue that if there are deviations from a linear time dependence of the gate potential, the modified Wigner function instead of the actual incoming Wigner function will be observed. 
We note that the Wigner distribution function has previously been found to be a useful concept in electron quantum optics \cite{ferraro_2013,glattli_2014,slava_peter}. Here it appears naturally as a quantum counterpart of the classical probability density.

Our manuscript is organized as follows: In Sec.\ \ref{sec:model} we give a precise definition of our system and of the approximations involved, as well as introduce the Wigner distribution function. Our main result, a quantum version of Eq.\ (\ref{eq:Q_semicl}), is presented and derived in Sec.\ \ref{sec:derivation}. In that Section we also discuss special cases, such as a gate voltage with a linear time dependence (for which Eq.\ (\ref{eq:Q_semicl}) is exact), a gate voltage with sharp edges, and the limit of a slowly varying gate voltage. Section \ref{sec:results} illustrates our results, by comparing the ``classical'' and ``quantum'' expressions for a few characteristic examples. In Section \ref{sec:experiment} we discuss the implication of our results for the analysis and improvement of electron tomography experiments. We conclude in Sec. \ref{sec:Conclusion}.
\section{Model}
\label{sec:model}

\begin{figure}
\center
\includegraphics[width=0.45\textwidth]{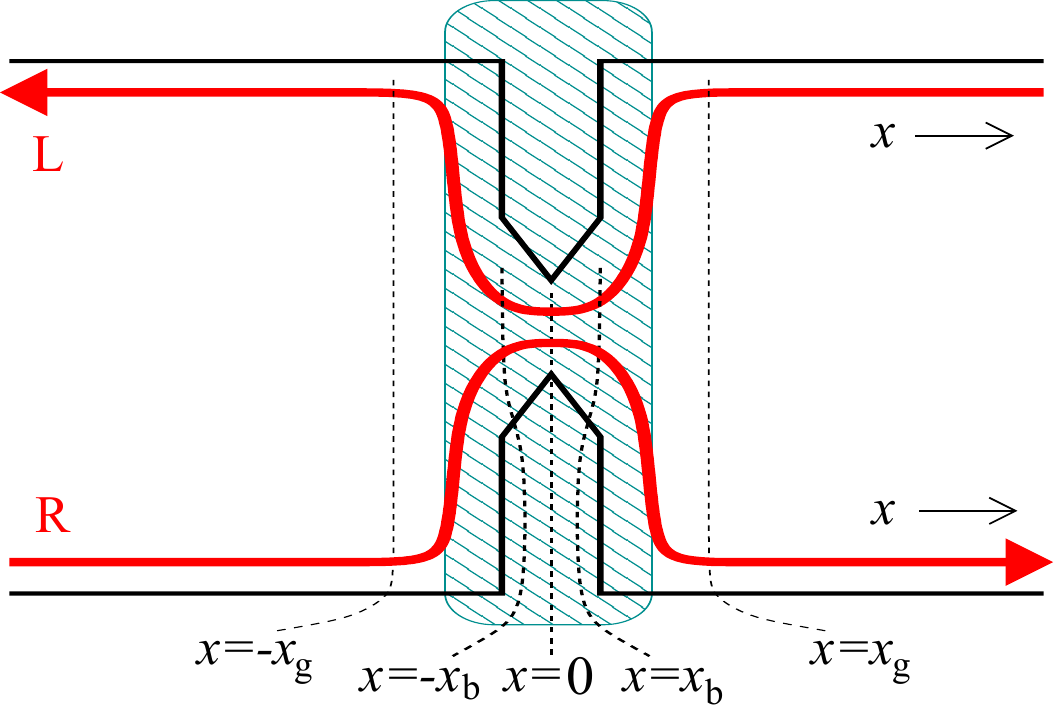}
\caption{\coloronline{(color online)}Constriction with two counterpropagating chiral edge modes. The coordinate $x$ measures the distance from the center of the constriction at $x=0$. Backscattering takes place within the region $-x_{\rm b} < x < x_{\rm b}$ only; the gate voltage $V(x,t)$ is spatially uniform and equal for both edge modes in this region. The gate voltage is zero for $x > x_{\rm g}$ and $x < -x_{\rm g}$.}
\label{fig:setup_exp2}
\end{figure}

We consider a constriction with two counterpropagating quantum Hall edge channels. The coordinate $x$ measures the distance along the edge, taking the center of the constriction as the origin, such that points to the left (right) of the constriction have negative (positive) $x$, see Fig. \ref{fig:setup_exp2}. The Hamiltonian for the two counterpropagating edge states reads
\begin{equation} \label{eq:hamchiral}
  \hat H = -i \hbar v \hat \sigma_3 \frac{\partial}{\partial x} + \hat H_{\rm bs} + \hat V(x,t),
\end{equation}
where $v$ is the electron velocity, $\hat \sigma_3$ is the Pauli matrix, and the $2 \times 2$ matrix structure refers to the two edges. The second term $\hat H_{\rm bs}$ describes backscattering in the constriction, and 
\begin{equation}
  \hat V(x,t) = \begin{pmatrix} V_{\rm R}(x,t) & 0 \\ 0 & V_{\rm L}(x,t) \end{pmatrix}
\end{equation}
is a gate potential. The backscattering Hamiltonian $H_{\rm bs}$ is characterized by its transmission probability
\begin{equation}
  T(E) = |\tau(E)|^2,
\end{equation}
in the absence of the gate voltage $V(x,t)$, where $\tau(E)$ is the transmission amplitude. 

Since backscattering takes place locally at the contact, the backscattering term has finite support which we define to be $|x| < x_{\rm b}$.
We furthermore assume that
\begin{equation}
  V_{\rm R}(x,t) = V_{\rm L}(x,t) = V(t)\ \mbox{for $|x| < x_{\rm b}$}.
  \label{eq:VLVR}
\end{equation}
This assumption is justified if backscattering exclusively takes place near the center of the constriction, which is where the two counterpropagating edge states are closest, and where the externally controlled part of the potential is spatially uniform. Although, strictly speaking, the condition \eqref{eq:VLVR} is stronger than the assumption that a (time-independent) shift of the gate voltage $V$ is equivalent to a shift in energy, the relation
 $T(E,V)=T(E-V)$ (which is central to the tomography experiment \cite{fletcher_2018}) does not hold for a generic scatterer if the condition \eqref{eq:VLVR} is not satisfied.
 
Finally, we set the gate potential to zero sufficiently far away from the constriction, 
\begin{equation}
  V_{\rm R}(x,t) = V_{\rm L}(x,t) = 0 \ \mbox{for $|x| > x_{\rm g}$}.
\end{equation}

We consider a wavepacket incident on the constriction, such that it reaches the center of the constriction for $t \approx 0$. For times $t \to -\infty$ this initial state is described by the wavefunction
\begin{equation}
  \psi_{\rm in}(x,t) = \varphi_{\rm in}(t-x/v) \begin{pmatrix} 1 \\ 0 \end{pmatrix}, \label{eq:psi_in}
\end{equation}
where the function $\varphi_{\rm in}(t)$ is peaked near $t=0$. The initial-state wavefunction is normalized such that
\begin{equation}
  \int_{-\infty}^{\infty} dt \, |\varphi_{\rm in}(t)|^2 = 1 \, .
\end{equation}
Similarly, for times $t \to \infty$ the wavepacket is described by the wavefunction
\begin{equation}
  \psi_{\rm out}(x,t) = \begin{pmatrix} \varphi_{\rm out}(t-x/v)  \\ \varphi_{\rm out}'(t+x/v)  \end{pmatrix}, \label{eq:psi_out}
\end{equation}
where $\varphi_{\rm out}(t-x/v)$ represents the transmitted part of the wave packet, and $\varphi_{\rm out}'(t+x/v) $ represents the reflected part.

Instead of the function $\varphi_{\rm in}(t)$ it is convenient to use the Wigner transform $\rho_{\rm in}(E,t)$, which is defined as
\begin{align}
  \rho_{\rm in}(E,t)=\frac{1}{\pi\hbar}\int\limits_{-\infty}^{+\infty} dt' e^{2iE t'/\hbar} \varphi_{\rm in}^*(t-t') \varphi_{\rm in}(t+t'). \label{eq:W_def}
\end{align}
The classical limit of the Wigner distribution function $\rho(E,t)$ is the joint probability density $p(E,t)$ of energy $E$ and time $t$ and corresponds to a Dirac delta distribution peaked around the classical trajectory $E(t)$. Correspondingly, the spread of the Wigner distribution around the classical trajectory is a manifestation of the quantumness of the wave packet~\cite{slava_peter}. The interpretation of $\rho(E,t)$ as a joint probability distribution is not rigorous, though, since the Wigner distribution can also take on negative values. Integrating $\rho(E,t)$ with respect to one of its arguments, however, one obtains a positive function, which is the probability density (marginal distribution) of the other argument, {\it i.e.},
\begin{align}
  p_E(E) &= \int\limits_{-\infty}^{+\infty}\rho(E,t) \, dt,\nonumber \\
  p_t(t) &= \int\limits_{-\infty}^{+\infty}\rho(E,t) \, dE.
\end{align}

\section{Transmitted charge\label{sec:derivation}}

\subsection{Derivation of the main result}

We now proceed with the calculation of the charge $Q$ transmitted through the constriction,
\begin{equation} \label{eq:Qdef}
  Q = e \int dt \, |\varphi_{\rm out}(t)|^2.
\end{equation}
For this calculation it is sufficient to consider the right-moving edge only. We drop the spinor notation of the previous Section, use the scalar wavefunction $\psi(x,t)$ to refer to the wavefunction component at the right-moving edge, and write $V(x,t)$ instead of $V_{\rm R}(x,t)$.

In the absence of the gate voltage, $V(x,t)=0$, the result can be easily expressed in terms of
the energy-dependent transmission probability $T(E)$,
\begin{equation}
  Q = e \int dE \, dt \, T(E) \, \rho_{\rm in}(E,t),
  \label{eq:QE}
\end{equation}
which is a special case of the classical equation (\ref{eq:Q_semicl}), with $p(E,t)$ replaced by $\rho_{\rm in}(E,t)$.

We now consider the general case of time- and energy-dependent scattering, specializing to the geometry described in Sec.\ \ref{sec:model}, for which the time dependence comes from the gate potential $V(x,t)$.

Since the potential $V(x,t) = 0$ for $x < -x_{\rm g}$, the initial-state solution (\ref{eq:psi_in}) is valid for all $x < -x_{\rm g}$. Similarly, the expression (\ref{eq:psi_out}) for the transmitted wavepacket is valid for all $x > x_{\rm g}$. In the first step of our calculation we solve the time-dependent Schr\"odinger equation to calculate the wavefunction at all positions $x \le -x_{\rm b}$ and $x \ge x_{\rm b}$,
\begin{align}
  \psi(x,t) &= \varphi_{\rm in}(t\!-\!x/v) \, e^{-\frac{i}{\hbar}\int\limits_{-\infty}^{t}dt' V[x-v(t-t'),t']}, & \mbox{$x \leq -x_{\rm b}$}, \nonumber \\
  \psi(x,t) &= \varphi_{\rm out}(t\!-\!x/v) \, e^{+\frac{i}{\hbar}\int\limits_{t}^{\infty}dt' V[x-v(t-t'),t']}, & \mbox{$x \geq  x_{\rm b}$}.
  \label{eq:psi1}
\end{align}
In a second step we perform the gauge transformation
\begin{align} \label{eq:gaugeTransform}
  \widetilde{\psi}(x,t)=\psi(x,t) \, e^{\frac{i}{\hbar}\int\limits_{-\infty}^t V(t') \, dt'},
\end{align}
where $V(t)$ is the value of the gate voltage at the center of the constriction, see Eq.\ (\ref{eq:VLVR}). For $-x_{\rm b} < x < x_{\rm b}$ the transformed wavefunction $\tilde \psi$ satisfies a Schr\"odinger equation without gate potential, so that
\begin{align}
  \widetilde{\psi}(x_{\rm b},t\!+\!x_{\rm b}/v)=\int \tau(t-t') \,
  \widetilde{\psi}(-x_{\rm b},t'\!-\!x_{\rm b}/v) \, dt',
  \label{eq:psitilde}
\end{align}
where
\begin{equation}
  \tau(t) = \frac{1}{2 \pi \hbar} \int dE \,  e^{-i E t/\hbar} \, \tau(E)
\end{equation}
is the Fourier transform of the constriction's transmission amplitude $\tau(E)$. The choice of the offsets in the time arguments in Eq.\ (\ref{eq:psitilde}) ensures that, for a perfectly transmitting constriction with $\tau(t) = \delta(t)$, $\widetilde \psi(x,t)$ is a function of $t-x/v$ only. 

Note that we do not need to invoke $\psi(x,t)$ explicitly  for $x \in [-x_b, +x_b]$ --- all the information about coherent dynamics inside the modulated scattering region is encoded in the matrix element $\tau(t)$ of the  time evolution operator which is time-translation invariant in the gauge expressed by Eq.~\eqref{eq:gaugeTransform}. This is valid as long as the modulation is spatially flat as expressed by condition \eqref{eq:VLVR}.
Matching Eqs.~(\ref{eq:psi1})  and \eqref{eq:psitilde} at $x=-x_b$ and $x=+x_b$, we find
\begin{equation}
  \varphi_{\rm out}(t) = \int dt' \, \tilde{\tau}(t,t') \, \varphi_{\text{in}}(t') \, ,
\end{equation}
with
\begin{align}
  \tilde \tau(t,t') =&\, e^{-\frac{i}{\hbar}\int\limits_{-\infty}^{t'}dt'' V[-v(t'-t''), t''] - \frac{i}{\hbar} \int\limits_{t'}^{t} dt'' V(t'')}
  \nonumber \\ & \mbox{} \times e^{- \frac{i}{\hbar} \int\limits_{t}^{\infty} dt'' V[-v(t-t''), t'']} \tau(t-t').
  \label{eq:tildetau}
\end{align}
Transmitted charge \eqref{eq:Qdef} can be written in the form that closely resembles the classical result (\ref{eq:Q_semicl}) and the limiting case (\ref{eq:QE}),
\begin{align}
  Q =&\,
  e \int \, dt \, dE \,  T(E) \, \tilde \rho(E,t),
  \label{eq:Q}
\end{align}
but with the function
\begin{align}
  \tilde \rho(E,t) =&\, \frac{1}{\pi \hbar} \int dt'
  \varphi_{\text{in}}(t\!+\!t') \varphi_{\text{in}}^{\ast}(t\!-\!t')
 \, e^{
  \frac{i}{\hbar}\!\!\int\limits_{t-{t'}}^{t+t'}\! dt_1 [E+\mathcal{V}(t_1)]}
  \label{eq:tilderho}
\end{align}
instead  of $p(E+V(t),t)$. Here
\begin{align} \label{eq:calVdef}
 \mathcal{V}(t) = \int_{-\infty}^{0} dx \left( \frac{\partial V(x,t')}{\partial x} \right)_{t' \to t + x/v} 
\end{align}
is the integral of the electric field along the electron's trajectory. 
The expressions \eqref{eq:Q}--\eqref{eq:calVdef} are the key results of this article.

The modified Wigner function $\tilde{\rho}(E,t)$ can be interpreted as the Wigner representation \eqref{eq:W_def}
of a modified incoming asymptotic state, $\tilde{\varphi}_{\text{in}}(t) =\varphi_{\text{in}}(t) \, e^{-i \int^{t} \mathcal{V}(t') d t'/\hbar}$. The effective gate potential 
$\mathcal{V}(t)$ represents the relevant modification of the applied gate potential $V(t)$ due to finite spatial extent of the gate.
Although $\tilde{\varphi}_{\text{in}}(t)$, and hence $\tilde{\rho}(E,t)$, does not in general represent the actual quantum state of the electron at any time instant, $\tilde{\rho}(E,t)$  can still  be measured as the outcome of a tomographic reconstruction, as we argue in Sec.\ \ref{sec:experiment}.

The main result \eqref{eq:Q} can easily be generalized to the case in which the
incoming state is not a pure state. In this case the Wigner function is
defined with the help of the single-particle density matrix $\hat \rho$ and the product
$\varphi_{\rm in}^*(t-t') \varphi_{\rm in}(t+t')$ in Eq.~\eqref{eq:Q} is
replaced by the matrix element $\langle t-t'|\hat \rho|t+t'\rangle$.

Our model Hamiltonian \eqref{eq:hamchiral} does not take into account decoherence during
the propagation along the chiral edge, or dispersion resulting from
(small) corrections to the linear kinetic energy term. %
Dispersion preserves the pure-state character of the incoming
wavepacket, whereas decoherence changes it into a mixed state, which
must be described using a density matrix (see the preceding remark).
If these processes are relevant (e.g., due to phonon emission \cite{Emary2019}), the modified Wigner function in Eq. \eqref{eq:tilderho}
must be calculated with respect to the Wigner function of the wavepacket
as it arrives at the constriction, {\em i.e.}, it should include the
decoherence and dispersion effects accumulated during the propagation
along the chiral edge. Dispersion or decoherence processes associated
with the (short) propagation inside the constriction region ({\em i.e.},
for $-x_{\rm g} < x < -x_{\rm b}$) are not included in our approach.
These can, however, be assumed to be small, since the size of the
constriction itself is typically much less than the distance between the
source and the constriction.

\subsection{Special cases\label{sec:SpecialCases}}
Below we consider a number of special cases of $V(x,t)$ and discuss the corresponding forms of $\mathcal{V}(t)$ and $\tilde{\rho}(E,t)$.

\paragraph{Time-independent potential.}

One verifies that for a time-independent potential  $\partial V(x,t)/\partial t =0$  we have $\mathcal{V}(t)=V$
and  
\begin{equation}
  \tilde \rho(E,t) = \rho_{\rm in}(E+V,t),
\end{equation}
where $V$ is the value of the spatially uniform potential in
the scattering region. We see that the time-independent $V$ just
changes the energy reference level in Eq.\ \eqref{eq:QE}, as expected
from gauge invariance.

\paragraph{Factorized time and coordinate dependence.}
If the time and coordinate dependence is such that $V(x,t)=u(x) \, V(t)$ where
the shape function $u(x)$ satisfies
\begin{align}
 u(x) = 
 \left\{ \begin{array}{ll}
 1 \, , & - x_{\rm g} < x < +x_{\rm g}, \\
 0 \, ,  &  x< -x_b \text{ or } x > +x_b 
  \end{array} 
  \right. \, 
\end{align}
then 
\begin{align} \label{eq:uconv}
  \mathcal{V}(t) = \int_{-x_g}^{-x_b} dx \,  V(t+x/v) \, \frac{d{u}}{d x}  \, .
\end{align}
Typically, $u(x)$ is expected to be a smooth sigmoid function, in which case the convolution
\eqref{eq:uconv} limits the sharpness of time-dependent features in $\mathcal{V}(t)$  compared to $V(t)$.

\paragraph{Gate potential with sharp edges.}
In the limit of a sharp edge, $-x_g \to -x_b$, Eq.~\eqref{eq:uconv} with $d{u}/{d x} \to \delta (x+x_g)$ is applicable, and
the correction of $V(t)$ becomes just the shift of the argument,
\begin{align} \label{eq:Vsharp}
  \mathcal{V}(t) = V(t - x_g /v) \, .
\end{align}  
We see that the modification of the wave-packet happens at the edge of the gate-affected region.

\paragraph{Linear time dependence.\label{sec:linear}}
The experiment of Ref.~\onlinecite{fletcher_2018} features a potential with a linear time dependence,
\begin{equation} \label{eq:linear}
  V(x,t) = V_0(x) + t \, V_1(x) \, ,
\end{equation}
which yields a linear $\mathcal{V}(t)$,
\begin{align}
  \mathcal{V}(t) =&\,
 V_0(0) - \frac{1}{v} \int^0_{-\infty} dx \, V_1(x) + V_1(0) \, t \, .
  \label{eq:tildeV}
\end{align}
In this case we find that the function $\tilde \rho(E,t)$ can be expressed directly in terms of the Wigner distribution $\rho_{\rm in}(E,t)$ of the incoming wavepacket,
\begin{align}
  \tilde \rho(E,t) =&\, \rho_{\rm in}[E+ \mathcal{V}(t),t] \, .
  \label{eq:linearRes}
\end{align}
Upon comparing Eqs.\ (\ref{eq:Q}) and (\ref{eq:linearRes}) with Eq.\ (\ref{eq:Q_semicl}) we conclude that for a potential with linear time dependence the quantum-mechanical theory reproduces the classical approximation (\ref{eq:Q_semicl}).

Note that $\mathcal{V}(t)$ is not in general equal to the value of the potential $V(x\!=\!0,t)=V(t)$ at the centre of the gate at the same time instant $t$, but rather at an earlier time corresponding to the effective position of the gate edge. We can define the latter explicitly by rewriting Eq.~\eqref{eq:tildeV} as $\mathcal{V}(t)=V(t+\tilde{x}_g/v)=V_0(0) + V_1(0)(t+\tilde{x}_g/v)$ where $\tilde{x}_g<0$ is  the center of mass of the accelerating  electric field distribution, $\tilde{x}_g = \int_{-\infty}^{0} x \, (dV_1/dx) \,  dx/V_1(0)$.
For a sharp edge, we have $\tilde{x}_g=-x_g$, in accord with Eq.~\eqref{eq:Vsharp}.

\paragraph{Slowly varying potential.}
If the rate of change of the gate voltage is slow compared to the velocity of the wave packet, it is reasonable to expand
\begin{align}
  V(x,t) = V_0(x) + V_1(x) t + \frac{1}{2} V_2(x) t^2 + \ldots \, .
\end{align}
Truncating after the quadratic term, this leads to the formal expression
\begin{align}
  \tilde{\rho}(E,t) =&\, e^{- \frac{1}{24} V_2(0) \hbar^2 \frac{\partial^3}{\partial E^3}} \rho_{\rm in}[E + \mathcal{V}(t),t] \, ,
  \label{eq:slow}
\end{align}
where
\begin{align}
  \mathcal{V}(t) = & \int_{-\infty}^{0} dx \left( \frac{\partial V(x,t')}{\partial x} \right)_{t' \to t + x/v} \nonumber \\
  =&\, V_0(0) - \int_{-\infty}^{0} dx \left[\frac{V_1(x)}{v} + \frac{V_2(x) x}{v^2} \right]  \nonumber \\
   +& \left(V_1(0) - \int_{-\infty}^{0} dx \frac{V_2(x)}{v}\right) t + \frac{1}{2} \, V_2(0) \, t^2 \, .
\end{align}
Equation (\ref{eq:slow}) contains the result (\ref{eq:linearRes}) for a gate potential with a linear time dependence and is a good starting point for an expansion in a small second derivative $V_2(x)$ of the time-dependent gate potential.

\paragraph{Linear time dependence superimposed on an arbitrary gate potential $V'(x,t)$.}
If on top of a linear time dependence the potential $V(x,t)$ also contains arbitrary additional terms $V'(x,t)$,
\begin{align}
V(x,t)=V_0(x) + t V_1(x) + V'(x,t) \, , \label{eq:V_prime}
\end{align}
the modified Wigner function $\tilde{\rho}(E,t)$ can be expressed as
\begin{align}
\tilde{\rho}(E,t) = \tilde{\rho}'(E+\mathcal{V}(t),t) \, , \label{eq:rho_prime}
\end{align}
where $\mathcal{V}(t)$ is given by Eq.\ \eqref{eq:tildeV} and $\tilde{\rho}'(E+\mathcal{V}(t),t)$ is calculated with respect to $V'(x,t)$ only.
The relevance of this result will be discussed in Sec.\ \ref{sec:experiment}. 
(Note that Eq.\ \eqref{eq:rho_prime} simplifies to Eq.\ \eqref{eq:linearRes} for the special case $V'(x,t)=0$.)

\section{Examples}\label{sec:results}
\begin{figure}
\center
\includegraphics[width=0.21\textwidth]{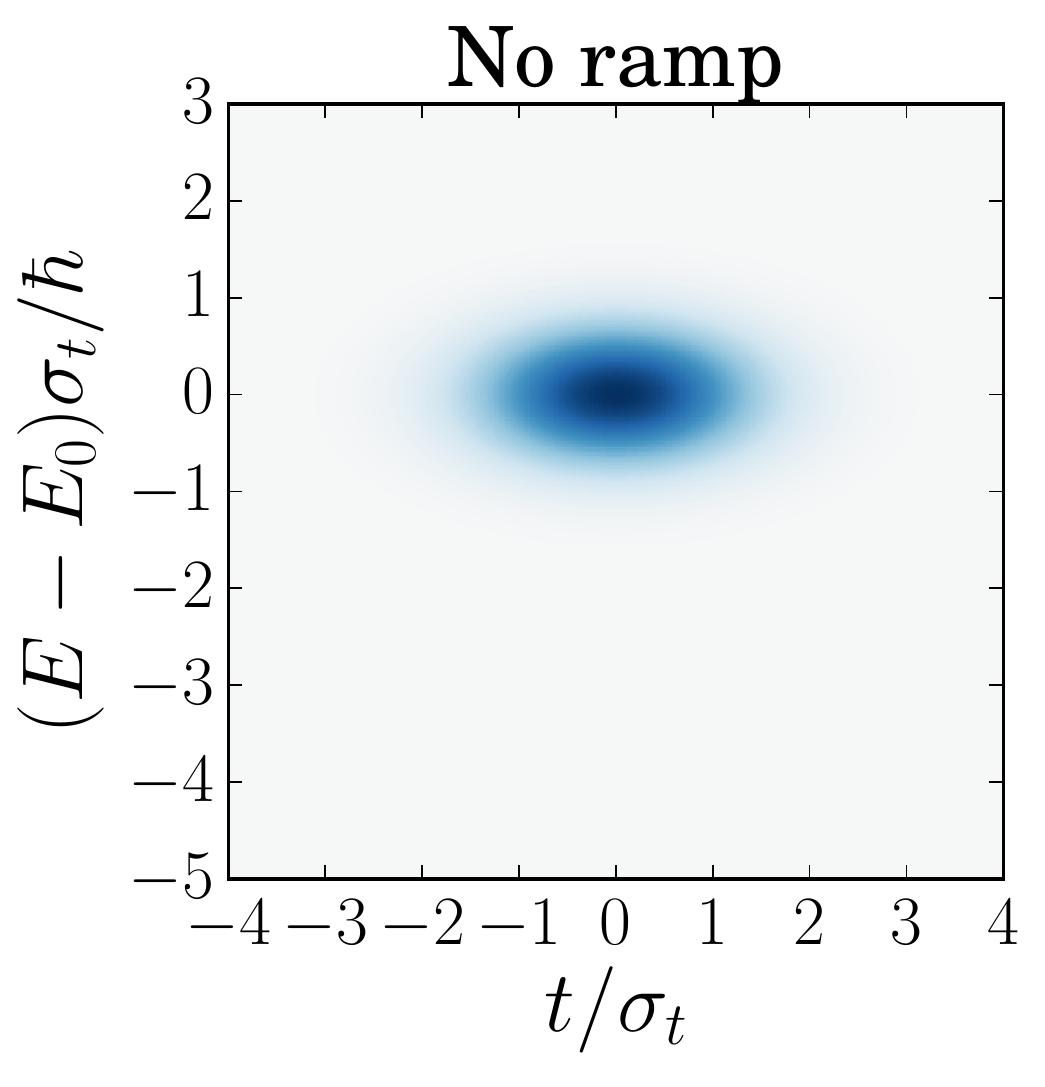}
\includegraphics[width=0.245\textwidth]{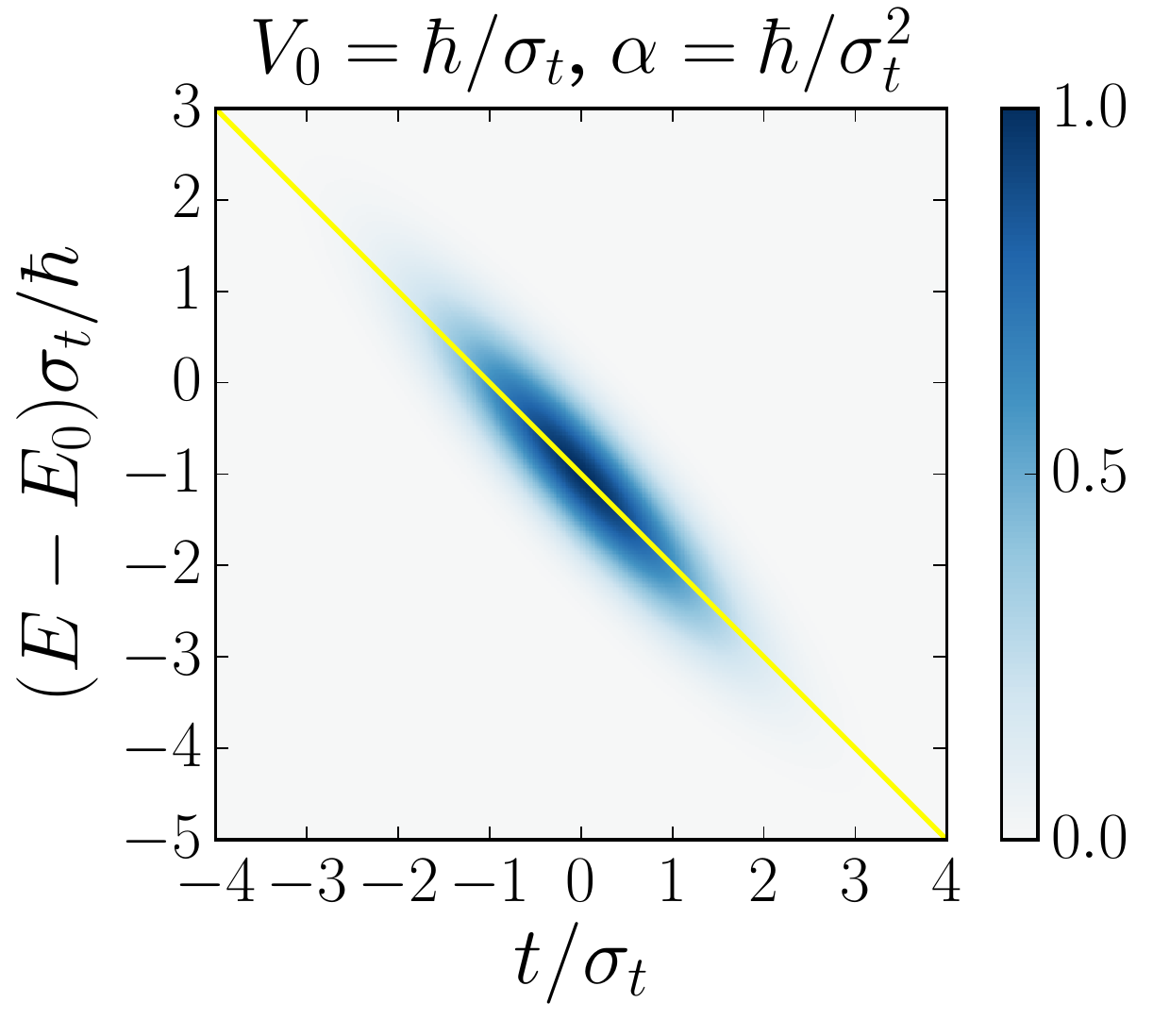} \\
\includegraphics[width=0.21\textwidth]{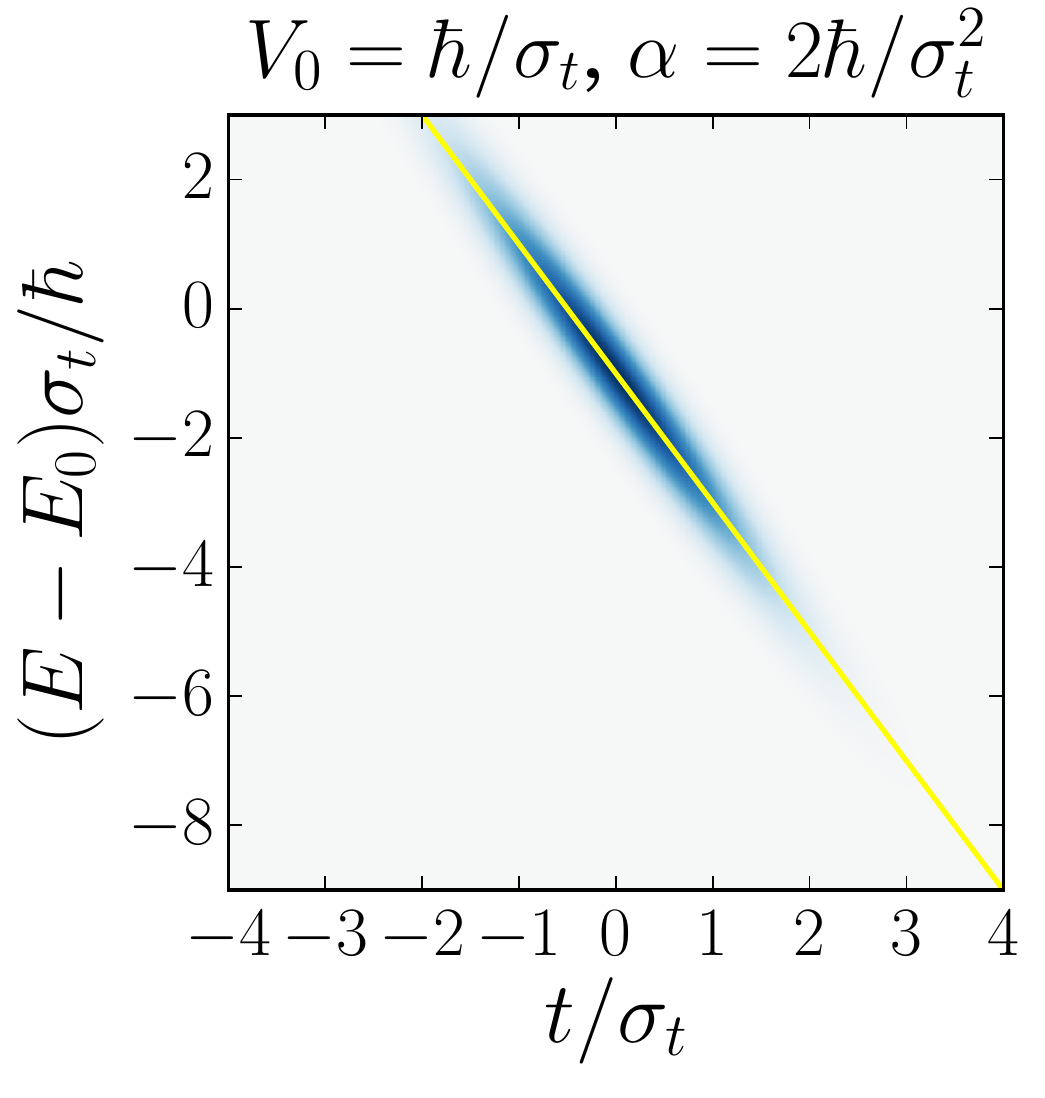}
\includegraphics[width=0.245\textwidth]{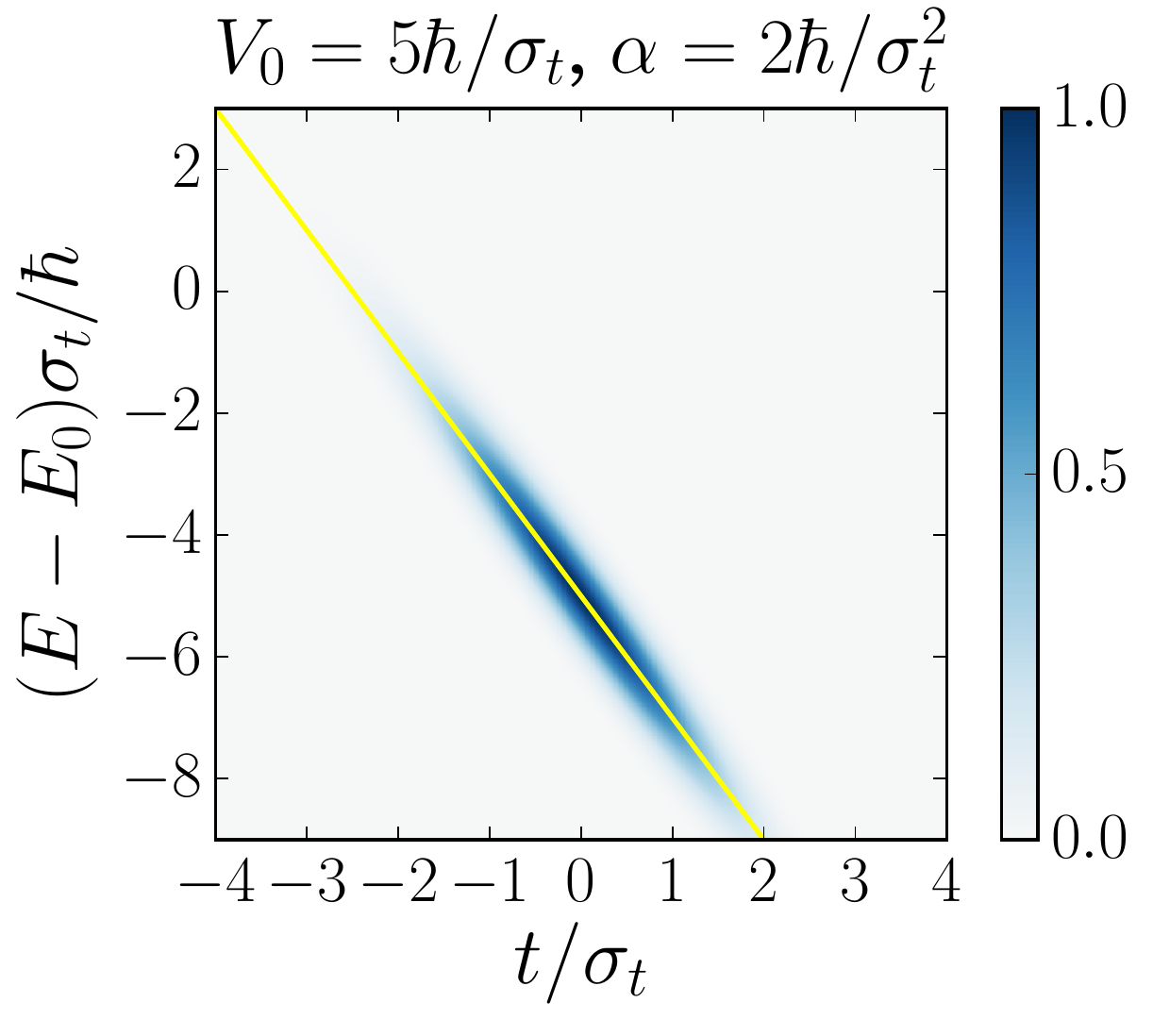} 
\caption{\coloronline{(color online)}Modified Wigner distribution in units of $1/\hbar$ of a Gaussian wave function after entering a gate with time-dependent potential $V(t) = \alpha t+V_0$ for different values of the ramp rate $\alpha$ and the offset $V_0$ as shown above. The upper left panel shows the Wigner distribution $\rho_{\rm in}(E,t)$ of the incoming wavepacket, the yellow line is $V(t)$.}
\label{fig:W_lin}
\end{figure}

In this Section we show explicit results for three examples: A gate voltage with linear time dependence, a gate voltage with an abrupt step-like time dependence, and a gate voltage with parabolic time dependence.
We take the spatial profile of the gate voltage $V(x,t)$ 
to be a spatially uniform potential with sharp edges at $x = \pm x_{\rm g}$. We also shift  the time origin by $-x_b/v$, so that the integral $\mathcal{V}(t)$ of the electric field can be replaced directly by the potential  $V(t)$ at the center of the constriction, see Eq.\ (\ref{eq:Vsharp}). For the incoming wavepacket we take the uncorrelated Gaussian form,
\begin{align}
  \varphi_{\rm in} (t) = \frac{1}{(2\pi \sigma_t^2)^{1/4}} e^{-\frac{i E_0 t}{\hbar}-\frac{t^2}{4\sigma_t^2}}. \label{eq:phi_in}
\end{align}
The scale $\sigma_t$ sets the characteristic width in the time domain. The energy of the wavepacket is centered around $E_0$, with fluctuations of order $\hbar/\sigma_t$. The Wigner distribution $\rho_{\rm in}(E,t)$ corresponding to the wavefunction (\ref{eq:phi_in}) is
\begin{equation} \label{eq:rho_in_gauss}
  \rho_{\rm in}(E,t) = \frac{1}{\pi \hbar} e^{-2 (E - E_0)^2 \sigma_t^2/\hbar^2 - t^2/(2 \sigma_t^2)}.
\end{equation}

For the examples in this and the following sections, we model the energy-dependence of the transmission probability through a static constriction as
\begin{align}
  T(E) & = \frac{1}{2} \left [ 1+ \erf \left (\frac{E}{\delta \sqrt{2}} \right)  \right ] \, ,
   \label{eq:T}
\end{align}
where $\delta$ gives the width of the energy window in which the transmission changes from $0$ to $1$ and the zero of energy is chosen to coincide with half-transmission point of the constriction. For the numerical calculations we will take the idealized limit $\delta \to 0$, corresponding to a point contact that perfectly selects the sign of the electron's energy.

\begin{figure}
\center
 \includegraphics[width=0.3\textwidth]{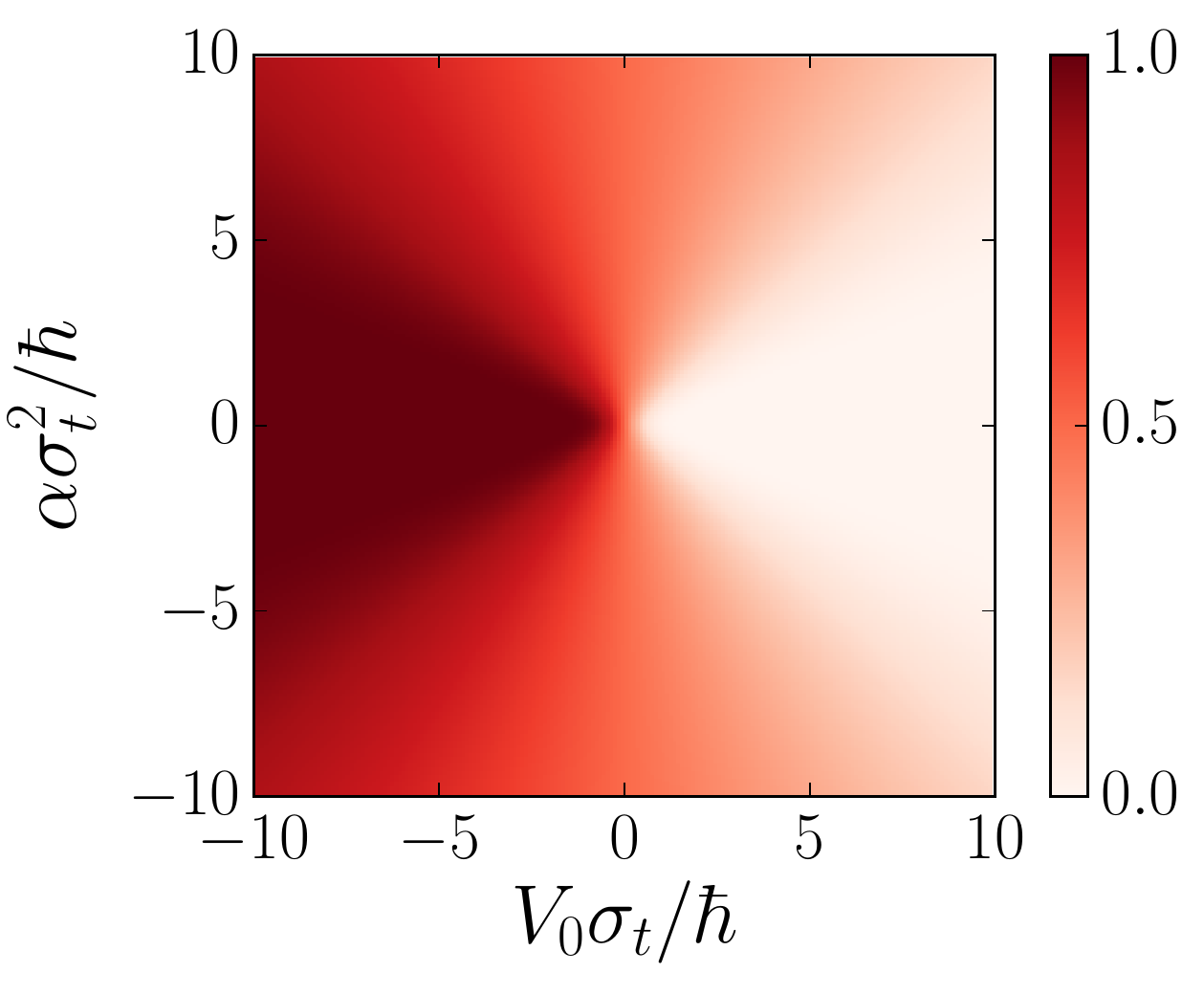}
\caption{\coloronline{(color online)}Transmitted charge through the constriction for the case of a linear voltage ramp, as a function of the ramp rate $\alpha$ and the offset $V_0$.}
\label{fig:charge_lin}
\end{figure}

\subsection{Linear ramp}\label{subs:linear}
We first consider the case of a gate voltage with a linear time dependence,
\begin{align}
  V(t)= \alpha \, t + V_0 \, . \label{eq:V_lin}
\end{align}
Gate voltages with different offsets $V_0$ are related by a delay time $t_{\rm d} = -V_0/\alpha$. For this case the modification of the Wigner distribution can be obtained from Eq.\ (\ref{eq:linearRes}), which gives:
\begin{align} \label{eq:W_lin}
\tilde{\rho}(E,t)=\frac{1}{\pi\hbar}e^{-2(E-E_0+V(t))^2 \sigma_t^2/\hbar^2 - t^2/2 \sigma_t^2}.
\end{align}
The modified Wigner distribution for different values of the ramp rate $\alpha$ and the offset $V_0$ is shown in Figure \ref{fig:W_lin}, along with the non-transformed Wigner distribution. In comparison to the original Wigner distribution $\rho_{\rm in}$, the modified distribution $\tilde \rho$ has a shifted center energy, determined by the offset $V_0$, and it is stretched along a straight line with slope $-\alpha$. Figure \ref{fig:charge_lin} shows the transmitted charge as a density plot in the case of a linear ramp as a function of the parameters $\alpha$ and $V_0$. 
This  transformation gives the basis for interpreting $p(E,t)$ in Eq.\ (\ref{eq:Q_semicl}) as the $\rho_{\text{in}}(E,t)$, as discussed in the discussed in Sec.\ \ref{sec:derivation}.

\subsection{Step-like ramp}
\begin{figure}
\center
\includegraphics[width=0.23\textwidth]{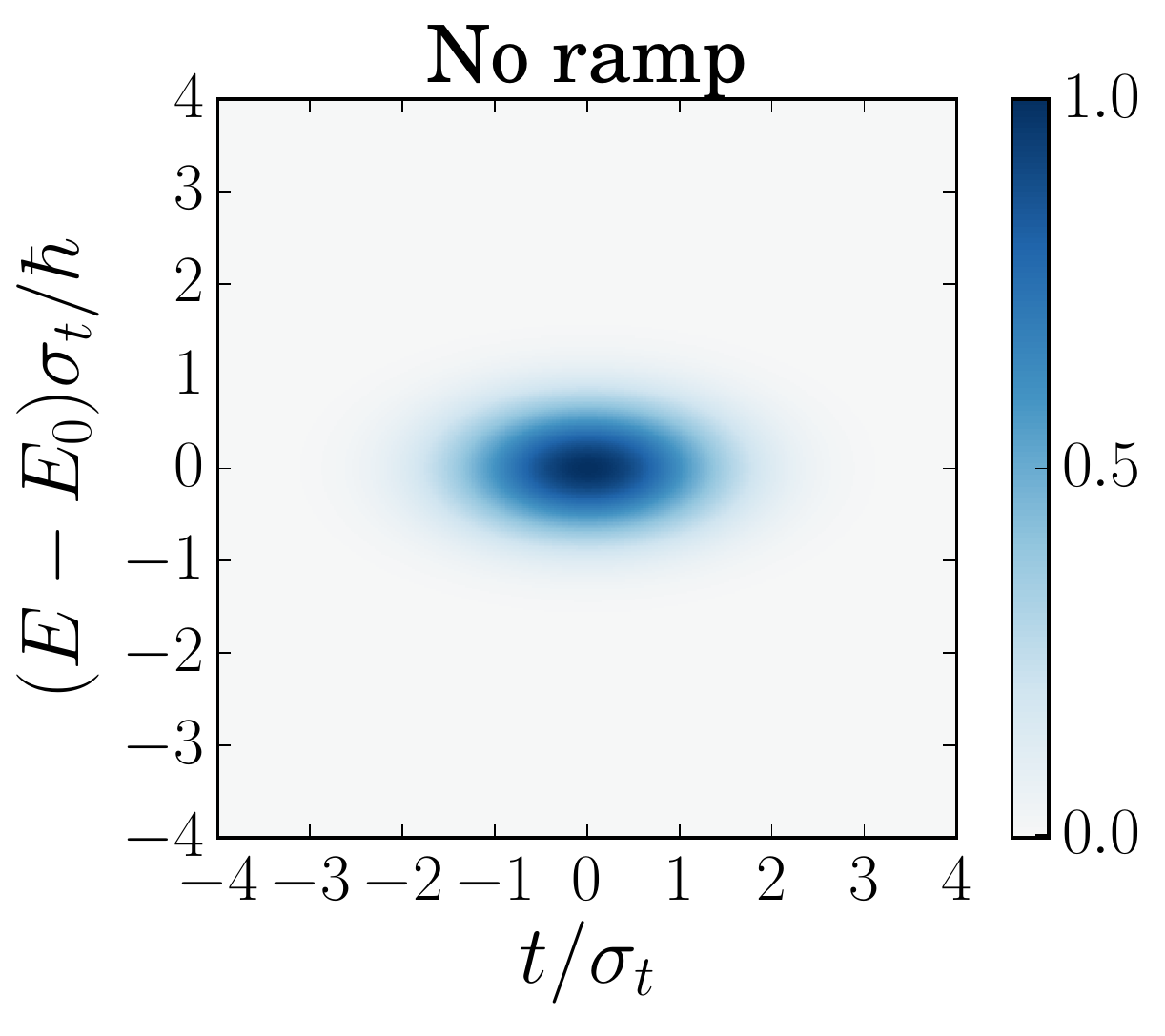}
\includegraphics[width=0.23\textwidth]{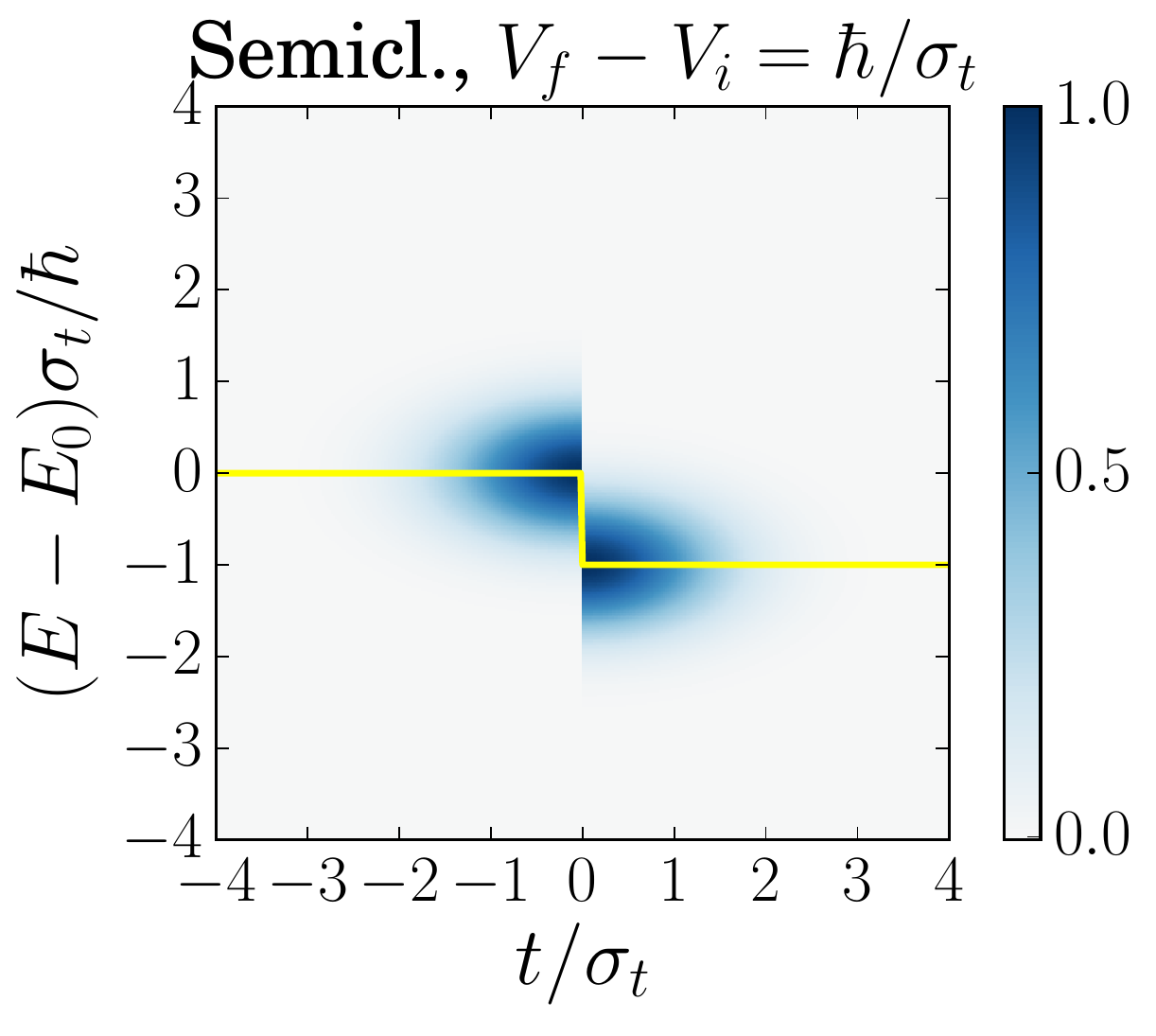} \\
\includegraphics[width=0.23\textwidth]{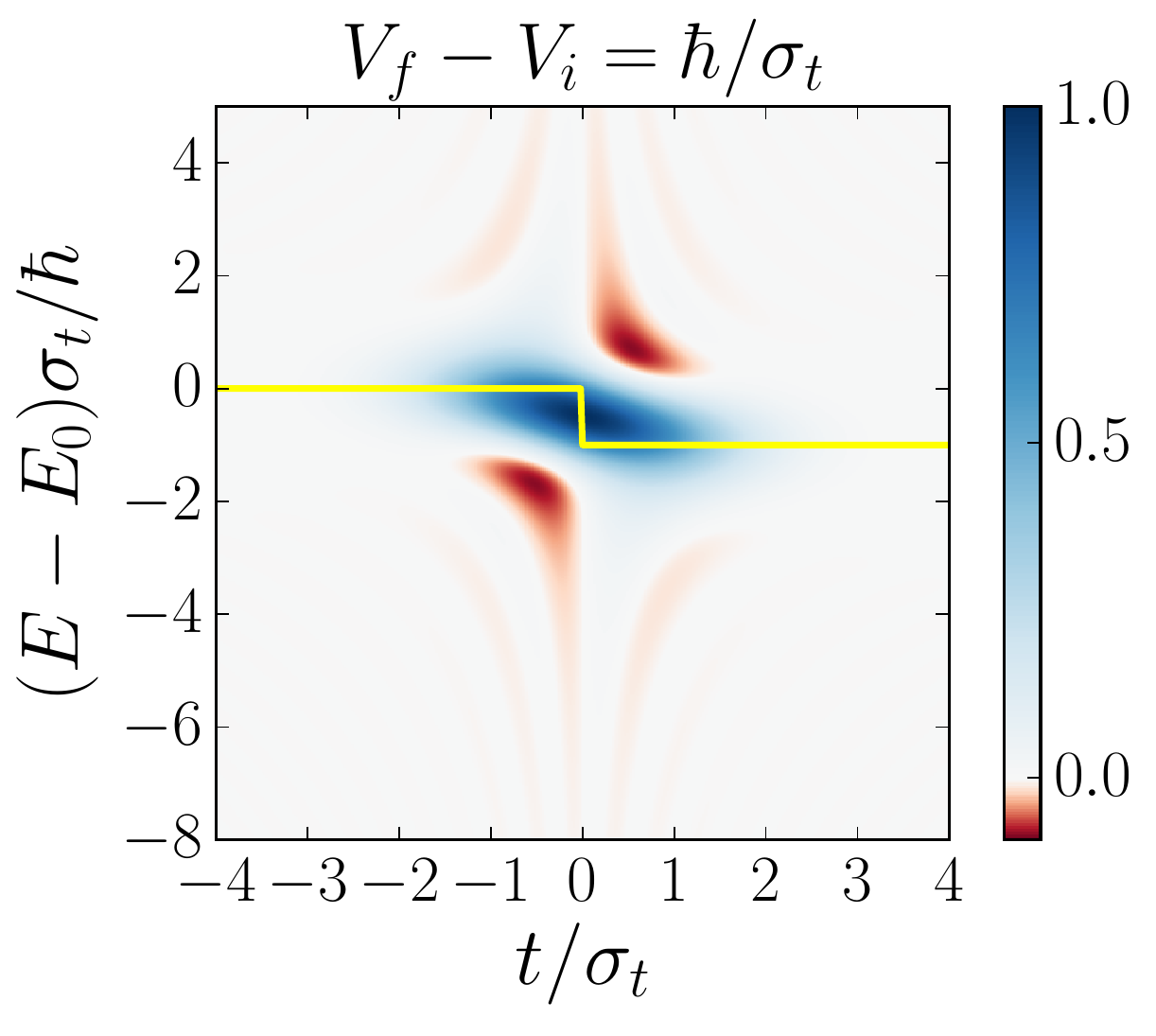}
\includegraphics[width=0.24\textwidth]{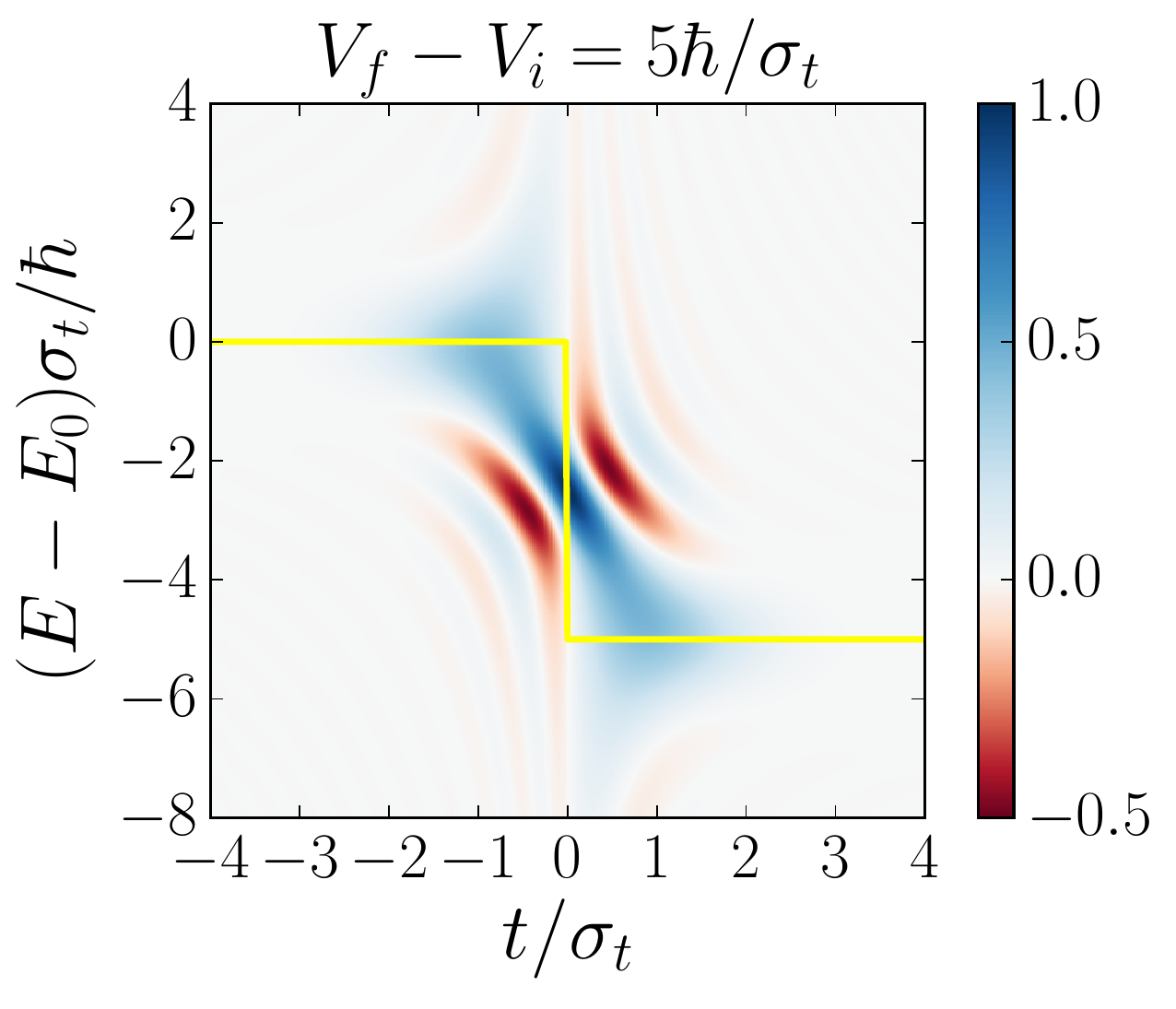} 
\caption{\coloronline{(color online)}Modified Wigner distribution in units of $1/\hbar$ of a Gaussian wave function after passing through a step-like ramp for different values of the barrier height $V_{\rm f}-V_{\rm i}$, compared to a non-transformed Wigner distribution (top left panel), as well as the semiclassical approximation (top right panel). We have set $V_{\rm i}=E_0$ and $t_0=0$.}
\label{fig:W_step}
\end{figure}

As mentioned in the introduction, performing the detection measurement with an energy-independent transmission probability that abruptly switches on or off allows one to measure the electron probability distribution as a function of time. 
Such an instantaneous switching on or off of the transmission function of the constriction requires an increase of the gate voltage $V(t)$ by an amount much larger than the energy uncertainty $\hbar/\sigma_t$ of the incoming wavepacket. Richer information about time and energy distribution can be obtained if the gate voltage jumps by a finite amount at time $t= t_0$,
\begin{equation}
  V(t) = \left\{ \begin{array}{ll} V_{\rm i} & \mbox{for $t < t_0$} \, , \\ 
  V_{\rm f} & \mbox{for $t > t_0$}\, . \end{array} \right  . \label{eq:V_step}
\end{equation}
For a Gaussian incoming wavepacket, the corresponding modified Wigner distribution can be calculated from Eqs.~\eqref{eq:tilderho} and \eqref{eq:Vsharp},
\begin{widetext}
\begin{align}
\tilde{\rho}(E,t)&=\frac{1}{\pi\hbar} e^{-2(E-E_0+V(t))^2\sigma_t^2/\hbar^2-\frac{t^2}{2\sigma_t^2}}\,
\re\left\{\erf\left[\frac{|t-t_0|}{\sqrt{2}\sigma_t}+\frac{i\sqrt{2}(E-E_0+V(t))\sigma_t}{\hbar}\right]\right\} \nonumber \\
&+\frac{1}{\pi\hbar}\re\Bigg\{e^{-2(E+{\bar V})^2\sigma_t^2/\hbar^2-\frac{t^2}{2\sigma_t^2}-i(V_{\rm f}-V_{\rm i})(t-t_0)/\hbar}
\erfc\left[ \frac{|t-t_0|}{\sqrt{2}\sigma_t}+\frac{i\sqrt{2}(E+{\bar V})}{\hbar} \right] \Bigg\}, \label{eq:rho_step}
\end{align}
\end{widetext}
where we abbreviated
\begin{equation}
  \bar V = \frac{1}{2}(V_{\rm f} + V_{\rm i}) - E_0 \, .
\end{equation}

A comparison between the initial-state Wigner function $\rho_{\rm in}(E,t)$, the modified Wigner function $\tilde \rho(E,t)$ for two different parameter choices, and the semiclassical expectation~\footnote{We call $\rho_{\rm in}(E+V(t),t)$  ``semiclassical'' because it can be interpreted as a quasi-probability density for electrons that have a well-defined energy $E+V(t)$ at every time instant $t$.} $\rho_{\rm in}(E+V(t),t)$ is shown Fig. \ref{fig:W_step}.
 Because of the abrupt time dependence at $t=t_0$, the semiclassical expectation is a poor approximation, as can be seen in the figure. In particular, the true modified distribution function is a continuous function of $E$ and $t$, whereas the semiclassical expectation has a discontinuous jump by $V_{\rm f} - V_{\rm i}$ at $t = t_0$. The two lower panels of the figure show that in contrast to a linear ramp, a step-like ramp transforms the Wigner distribution such that it assumes negative values.

\begin{figure}
\center
\includegraphics[width=0.235\textwidth]{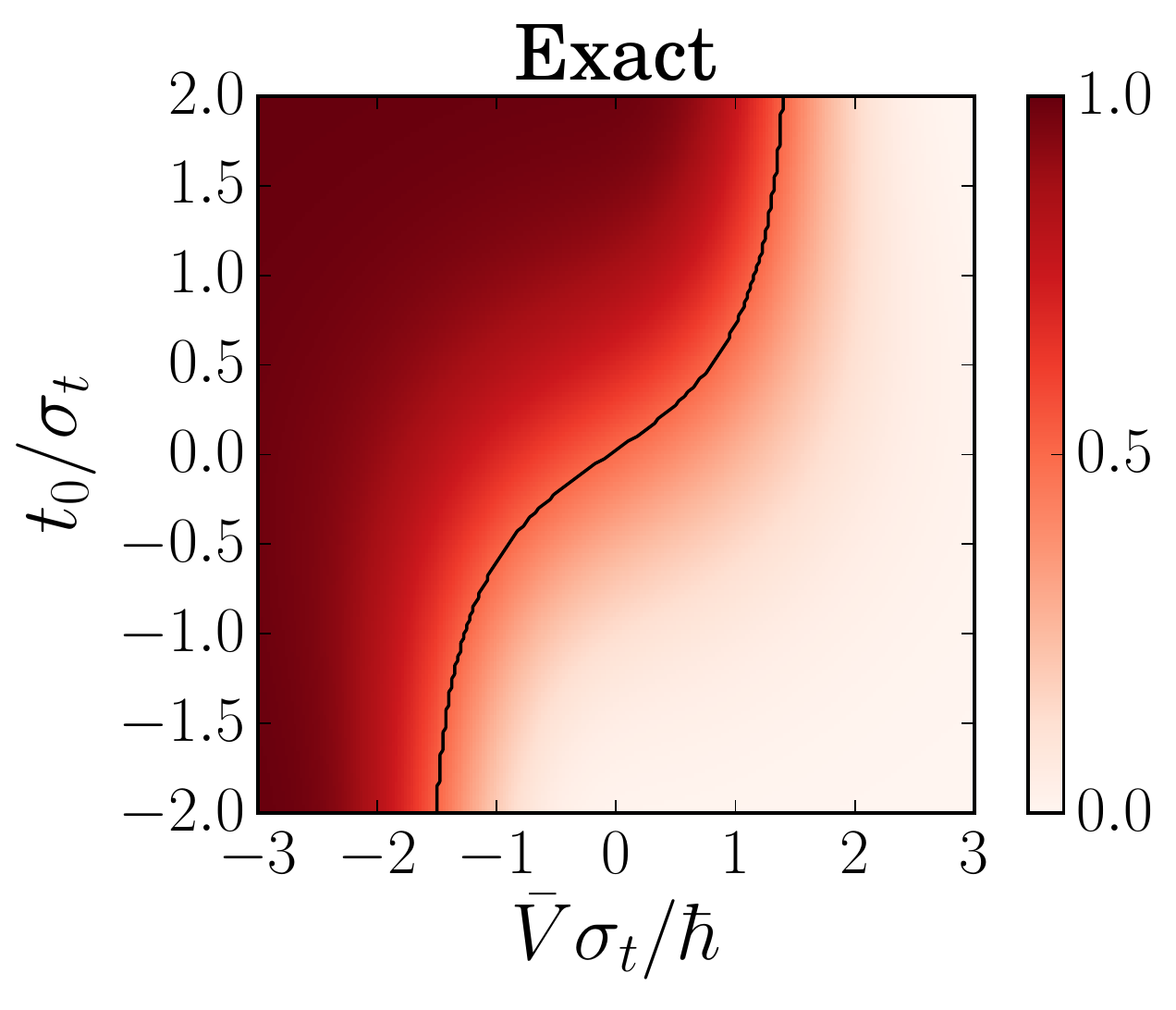}
\includegraphics[width=0.235\textwidth]{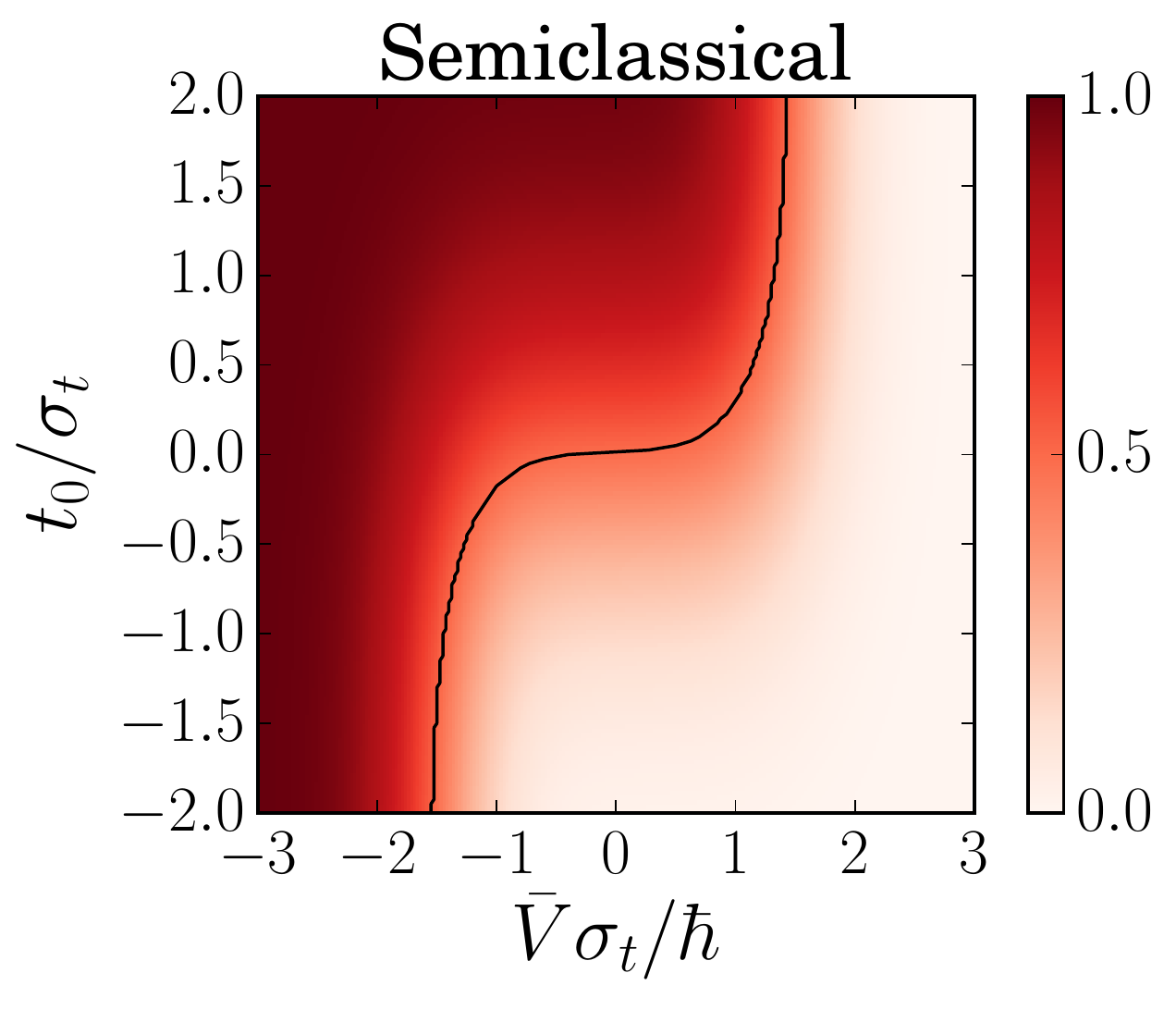}
\includegraphics[width=0.24\textwidth]{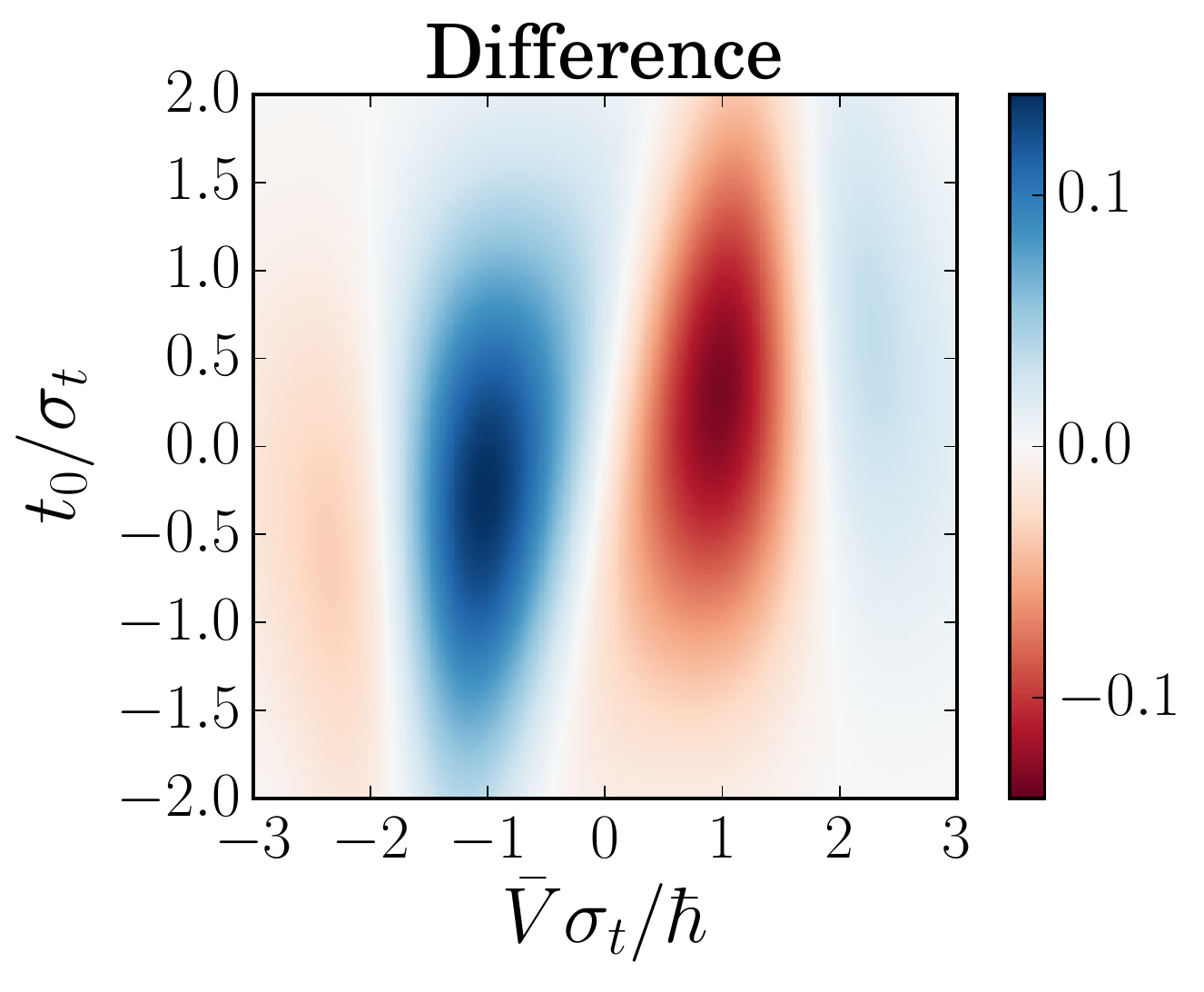}
\caption{\coloronline{(color online)} Transmitted charge for the case of a step-like ramp as a function of $\bar{V}=(V_{\rm f}+V_{\rm i})/2-E_0$ and the switching time $t_0$. The height of the step $V_{\rm f}-V_{\rm i}$ is kept fixed and equal to $3 \hbar/\sigma_t$. The top left panel shows the exact result; the top right panel shows the semiclassical approximation [(\ref{eq:Q_semicl}) with $p \to \rho_{\text{in}}$]. The black curve in the two top panels shows the $Q = e/2$ median. The bottom panel shows the difference between the exact and the semiclassical result.}
\label{fig:charge_step_Vaver}
\end{figure}

In Fig.\ \ref{fig:charge_step_Vaver} we show the transmitted charge at a fixed step height $V_{\rm f} - V_{\rm i}=3 \hbar/\sigma_t$ as a function of the center potential $\bar V$ and the switch time $t_0$. The top left panel shows the exact result (\ref{eq:Q}), and the top right panel the semiclassical approximation (\ref{eq:Q_semicl}). Their difference is shown in the bottom panel. As expected, if $|\bar V|$ is sufficiently large, the transmitted charge is independent of $t_0$ and approaches $0$ or $1$ (corresponding to maximum possible transmittance). For intermediate values of $\bar V$, there is a transition from $0$ to $1$ transmitted charge as $t_0$ goes from large negative to large positive values. Although the exact and the semiclassical results both reproduce the correct limits for large $|\bar V|$ and $|t_0|$, the behavior for small $\bar V$ and $t_0$ shows qualitative differences (such as shape of the median $Q = e/2$ curve), as well as quantitative differences (more than $0.1 e$, see bottom panel).

We can check explicitly that a large-amplitude sudden step in $V(t)$ will sample the time distribution. For this we need to prove that for  $-V_{\text{i}}, V_{\text{f}} \gg \hbar/\sigma_t, \delta$, finite $E_0$, and $ \tilde{\rho}(E,t)$ given by Eq.~\eqref{eq:rho_step}, the transmitted charge
\begin{align} \label{eq:Qcut}
 Q & = \int_{-\infty}^{\infty} d E \, \int_{-\infty}^{\infty} dt \, T(E) \, \tilde{\rho}(E,t) 
 \end{align}
equals to 
$Q=\int_{-\infty}^{\infty} d E \, \int_{-\infty}^{t_0} dt \,  \rho_{\text{in}}(E,t)$.
Of the two terms in Eq.~\eqref{eq:rho_step}, the contribution of the second one to the time integral in Eq.~\eqref{eq:Qcut} vanishes the limit of large $V_{\text{f}}-V_{\text{i}}$ due to fast oscillations, while the first term contributes only at $t<t_0$, when integrand is non-zero in the vicinity of $E\approx -V_{\text{i}}$. Taking  the limit $V_{\text{i}} \to - \infty$, this gives
\begin{align} \label{eq:Qalmost}
   Q = \int^{t_0}_{-\infty} dt \int_{-\infty}^{\infty} d E'  \rho_{\text{in}}(E',t)  \{ 1 - \re  
    \, e^{z^2} w(z) ]\} \, ,
\end{align} 
where $z=E'\sqrt{2} \sigma_t /\hbar +i (t_0-t)/ (\sqrt{2}\sigma_t) $ and $w(z)$ is Faddeeva function $w(z) = (i /\pi) \int e^{-\xi^2}/(z-\xi) \, d\xi$ ($\im z>0$).
The first term in curly brackets in Eq.~\eqref{eq:Qalmost} gives the expected result.
Straightforward integration confirms that the second term, proportional to $w(z)$, contributes zero.

\subsection{Parabolic ramp}

The third example is the case of a parabolic ramp, which may serve as an approximation of a periodic (sinusoidal) gate potential near the maximum or minimum of the potential. We consider a gate potential of the form
\begin{align}\label{eq:par_pot}
V(t)=V_0- \frac{1}{2} W(t-t_0)^2.
\end{align}
The modified Wigner function is
\begin{align}\label{eq:rho_par}
\tilde{\rho}(E,t)
 =&\, \frac{\sqrt{2/\pi}}{\sigma_t \sqrt[3]{\left| W\right|\hbar^2}}
e^{-\frac{E-E_0+V(t)}{W\sigma_t^2}-\frac{t^2}{2\sigma_t^2}+\frac{\hbar^2}{12 W^2\sigma_t^6}} \\ \nonumber & \times
\text{Ai}\left[-\frac{2}{\sigma_t\sqrt[3]{W \hbar }}\left(E-E_0+V(t)-\frac{\hbar}{8 W\sigma_t^3}\right)\right],
\end{align}
where $\text{Ai}(z)$ is the Airy function of the first kind.
\begin{figure}
\center
\includegraphics[width=0.49\textwidth]{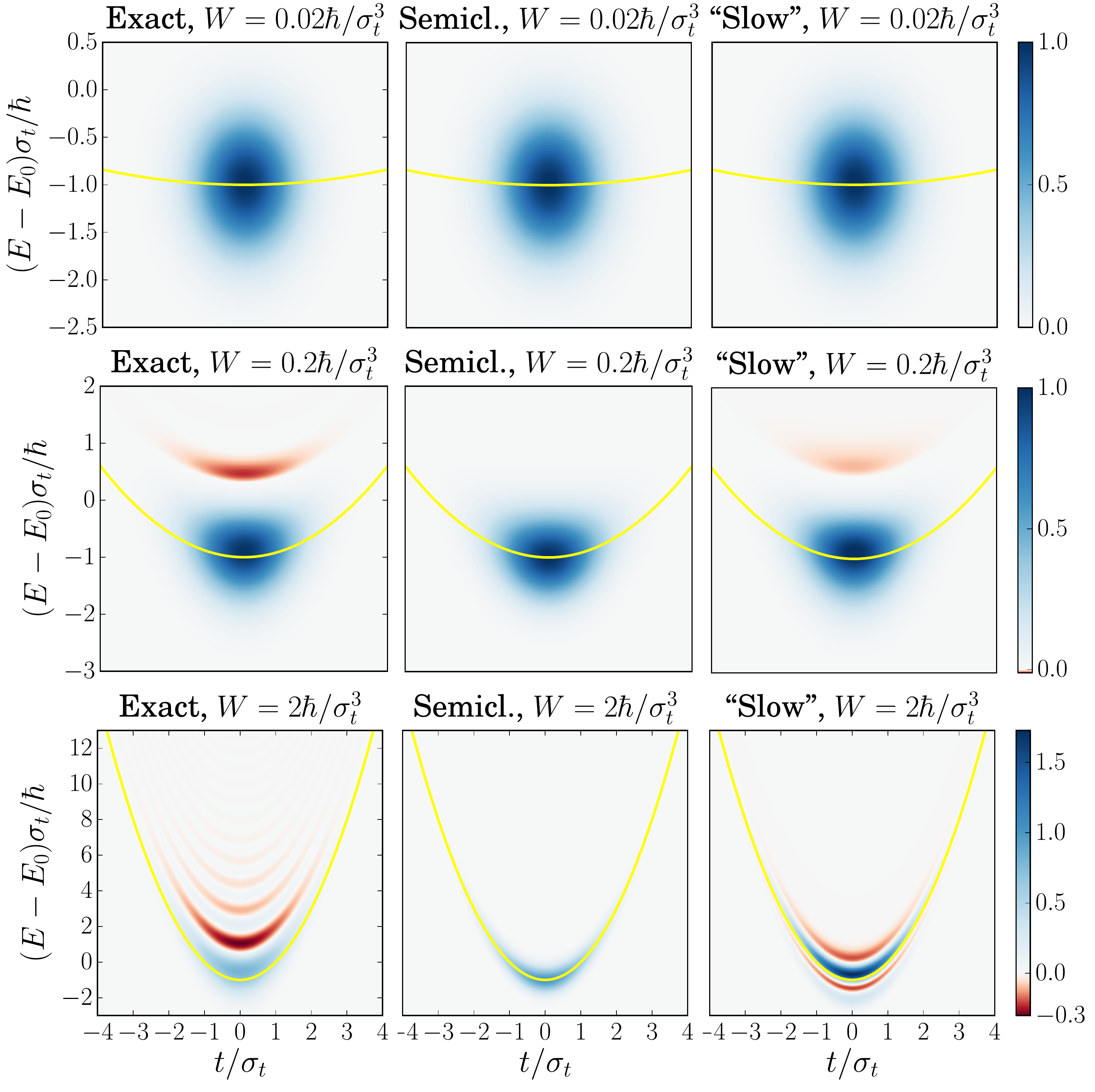}
\caption{\coloronline{(color online)} Modified Wigner distribution $\tilde \rho$ in units of $1/\hbar$ for a Gaussian wavepacket incident on the constriction. The gate potential $V(t)$ is taken to have the parabolic time dependence (\ref{eq:par_pot}) with $W= 0.02 \hbar/\sigma_t^3$ (top), $W = 0.2 \hbar/\sigma_t^3$ (center), and $W = 2 \hbar/\sigma_t^3$ (bottom). We have set $V_0=E_0+\hbar/\sigma_t$ and $t_0=0$. The left column shows the full quantum-mechanical result (\ref{eq:Q}), the center column shows the semiclassical approximation (\ref{eq:linearRes}), and right column shows the slow-potential approximation, obtained by truncating the exponential in Eq.\ (\ref{eq:slow}) at first order in $W$.}
\label{fig:W_par}
\end{figure}

Figure \ref{fig:W_par} contains a comparison between the exact result for the modified Wigner function, the semiclassical approximation (\ref{eq:linearRes}) and a slow-potential approximation obtained by truncating the exponential in Eq.\ (\ref{eq:slow}) at first order in the second derivative of the potential, for different values of $W$ and for $V_0 = E_0 + \hbar/\sigma_t$, $t_0 = 0$. The figure confirms that both the semiclassical and the slow-potential approximation are good approximations for the modified Wigner function $\tilde \rho(E,t)$ if $|W| \ll \hbar/\sigma_t^3$, whereby the slow-potential approximation also faithfully reproduces some of the fringes at larger values of $W$, which are absent from the semiclassical approximation.

\begin{figure}
\center
\includegraphics[width=0.49\textwidth]{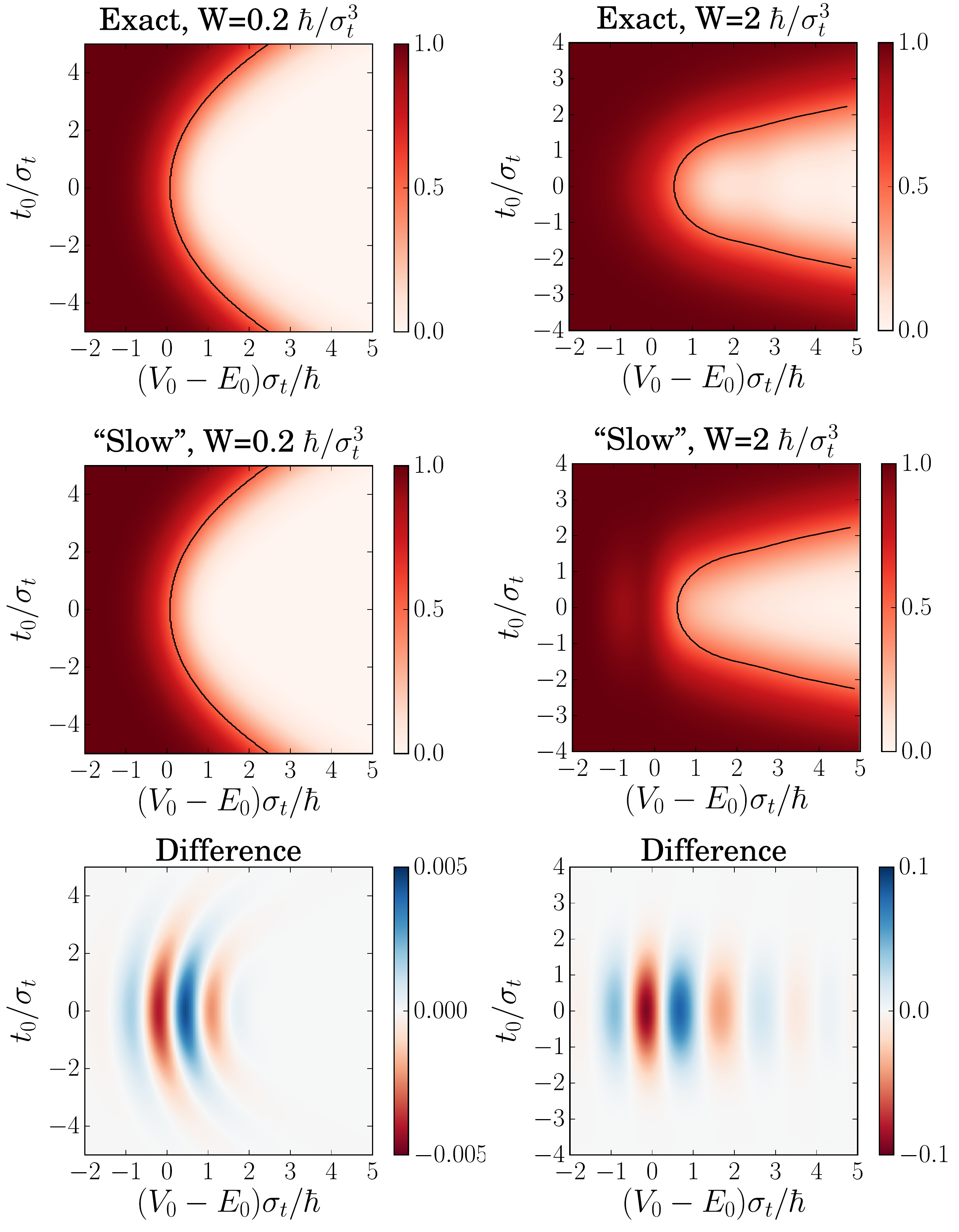}
\caption{\coloronline{(color online)} Transmitted charge $Q$ for a Gaussian wavepacket incident on a constriction subject to a gate potential with the parabolic time dependence (\ref{eq:par_pot}) with $W = 0.2 \hbar/\sigma_t^3$ (left) and $W = 2 \hbar/\sigma_t^3$ (right). The top row is obtained from the full quantum-mechanical expression Eq.\ (\ref{eq:Q}), the center row uses the slow-potential approximation (\ref{eq:slow}) with the exponential truncated after first order in the second derivative of the potential. The bottom row shows the difference between the slow-potential approximation and the full quantum-mechanical result.}
\label{fig:charge_par}
\end{figure}

In Fig. \ref{fig:charge_par} we show the transmitted charge for two different values of the the curvature parameter $W$, with separate panels for the exact result, the semiclassical approximation, and the slow-potential approximation. Their difference is shown in the bottom panels. The transmitted charge goes from zero for values of $V_0-E_0$ far below the line $V_0-E_0=\frac{1}{2}Wt_0^2$ to $1$ for values of $V_0-E_0$ far above this line. The differences between the exact and the slow-potential solution grows with $W$, remaining below 1\% for $W=0.2\hbar/\sigma_t^3$ and going up to 10\% for $W=2 \hbar\sigma_t^2$.

\section{Extension to non-chiral scattering\label{sec:numerics}}
The results of Section \ref{sec:derivation} and the examples discussed so far are exact for the model defined in Section \ref{sec:model} which includes a strict separation of the scattering and the acceleration/deceleration regions, as well as linear dispersion.
Here we demonstrate that our main results are robust against relaxing these assumptions, and show that Eqs.~\eqref{eq:Q}--\eqref{eq:calVdef} can provide a rather 
accurate approximation for scattering by a time-dependent one-dimensional potential controlled by a single gate even in the absence of a back-scattering-suppressing  magnetic field. 

\begin{figure} 
\center
\includegraphics[width=0.95\columnwidth]{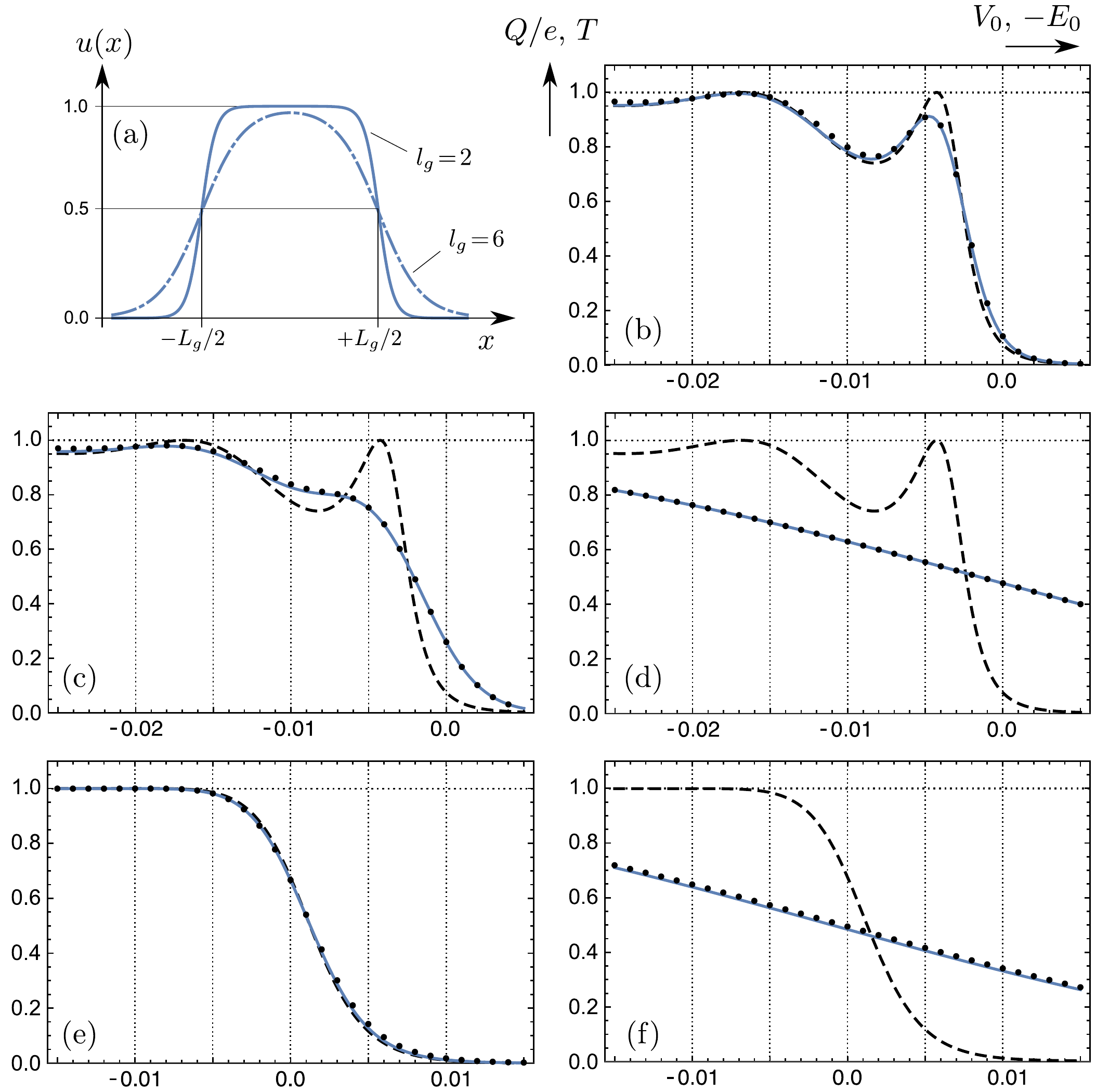}
\caption{(a) The real-space shape of the gate potential $u(x)$, for  $l_g=2$ (continuous line) versus $l_g=6$ (dash-dotted line) at $L_g=50$.
  (b)-(f) Comparison of numerical solution for time-dependent scattering in the non-chiral model defined by Eqs.~\eqref{eq:ham1D} (dots) and the formula \eqref{eq:Qapprox} (continuous line).  The graphs show the average transmitted charge $Q/e$ as function of $V_0$ ;
  the dashed line in the background is the static transmission probability $T(E_0+V_c)$, as function of an incoming plane-wave energy, $E_0$.
   Parameters of the scattering potential:
(b,c,d): $l_g=2$, (e,f): $l_g=6$, (b,e): $\alpha=0$, (c): $\alpha=5 \cdot 10^{-6}$, (d): $\alpha=5 \cdot 10^{-5}$, (f); $\alpha=-5 \cdot 10^{-5}$.
 All data are computed with $V_c=E(k_0)=0.062$, $L_g=50$ and $\sigma_E =0.001$. \label{fig:results1D}}
\end{figure}

In this section we compare analytical results of Section~\ref{sec:derivation} to a direct numerical  solution of the time-dependent Schr\"odinger equation for the following Hamiltonian:
\begin{align} \label{eq:ham1D}
\hat{H}_{\text{1D}}  = & -\frac{\hbar^2}{2 m^{\ast}} \frac{\partial^2}{\partial x^2} + V_c \, u(x) + V(t)\,  u(x) \, ,
\end{align}
\begin{align}u(x)  = & \frac{1}{1+\exp[(x-L_g/2)/l_g] } 
 - \frac{1}{1+\exp[(x+L_g/2)/l_g] }  \, . \label{eq:potential}
\end{align}
The second term and the third terms in Eq.~\eqref{eq:ham1D} mimic the constriction part $\hat{H}_{\text{bs}}$ and the gate part $V(x,t)=V(t) u(x)$ of the Hamiltonian \eqref{eq:hamchiral},  respectively. 
For $L_g \gg l_g$ the potential form factor $u(x)$ is a mesa of height 1 with a flat plateaux part for $-L_g /2 < x < +L_g/2$, and rounded edges of characteristic width $l_g$ [see Fig.~\ref{fig:results1D}(a)]. This approximates the condition of spatially flat modulation. In the absence of modulation, $V(t)=0$, the barrier height $V_c$ determines the transmission energy threshold. We emphasize that \eqref{eq:ham1D} defines a model
different to that of Section \ref{sec:model}, because both potential terms  contribute to back-scattering between the left- and the right-movers; an approximate equivalence is expected only if $|V(t)| \ll V_c$ during scattering.
In the comparison below, we consider  linear modulation only, $V(t)=V_0 +\alpha \, t$ as in Section~\ref{sec:results}.A, cf.~Eq.~\eqref{eq:V_lin}. 

For the numerical solution, the $x$-coordinate is discretised on a regular mesh with lattice spacing $a$ and total number of sites $N \approx 16 000$,
the  Hamiltonian \eqref{eq:ham1D} is approximated by a nearest-neighbour tight-binding Hamiltonian with hopping amplitude $-J$.
The latter is fixed by matching the parabolic approximation at the bottom of the tight-binding band to the kinetic energy part of Eq.~\eqref{eq:ham1D}, i.e.\ $E(k)=2 J (1 - \cos ka) = \hbar^2 k^2/(2 m^{\ast}) + O(k^4 a^4)$. 
Units of energy, length, and time are fixed by setting $J=1$, $a=1$ and $\hbar=1$, which gives $m^{\ast} =1/2$.

We follow propagation of the Gaussian wave-packet
\begin{align}
   \mathrm{\psi}_0(x) & \propto e^{i k_0 x} e^{-(\delta k)^2 x^2}  \, , \label{eq:GaussianNonChiral}
\end{align}
which corresponds to approximately normal energy distribution of width $\sigma_E= \hbar^2 k_0 \, \delta k/m^{\ast}$ centered at $E(k_0)$.


Split-step Fourier method \cite{Feit1982,Bellentani2018} is used to implement time evolution generated by $\hat{H}_{\text{1D}}(t)$.
The initial state $\psi_{\text{in}}(x,t_0)=e^{+ i \hat{H}_0 t_0} \psi_0(x) $, where $\hat{H}_0=-\hbar^2 (\partial /\partial x)^2/(2 m^{\ast})$, and time $t_0<0$ are chosen such that
$\psi_{\text{in}}(x,t_0)$ is centred at least 10 standard deviations from the edge $\tilde{x}_g=-L_g/2$.
This initial wave-packet is the propagated with full $\hat{H}_{\text{1D}}(t)$ from $t=t_0$ up to time $t=+1.5 |t_0|$, and the transmitted fraction is computed by projecting the final state wave-function onto the subspace of right-movers [wave numbers $0< k < \pi/(2 a)$].


We set the working point for comparison to the linearised model at $V_c=E(k_c)$ with $k_c a=0.25$ chosen so that the tight-binding approximation of the parabolic dispersion remains sufficiently accurate while 
the corresponding $E(k_c)=0.062$ is large enough to explore the regime $|V_0|, \sigma_E \ll V_c$.

The numerical correspondence between the Gaussian wave-packets in the two models, Eq.~\eqref{eq:phi_in} and Eq.~\eqref{eq:GaussianNonChiral}, respectively, is established by $\psi_0(x) = \varphi_{\text{in}}(-x/v) e^{+i k_0 x-i E_0 \, x /( \hbar v)} \approx  \varphi_{\text{in}}(-x/v)  e^{i k_c x}$,
with the velocity $v= \hbar^{-1} \partial E(k)/\partial k \rvert_{k=k_c} \approx \hbar k_c/m^{\ast}=0.5$ and the widths relation $\sigma_t \, \sigma_E = \hbar/2$. Here $e^{i k_c x}$ the carrier plane wave for which the amplitude distribution $\psi(x,t)$ [Eq.~\eqref{eq:psi_in}] in the linearized model \eqref{eq:hamchiral} acts as an envelope. 
Changing $k_0$ around $k_c$ in Eq.~\eqref{eq:GaussianNonChiral} corresponds to changing  $E_0= E(k_0)-E(k_c)\approx \hbar v (k_0-k_c)$ around $0$ in Eq.~\eqref{eq:phi_in}.

The modified Wigner function \eqref{eq:W_lin} for the  Gaussian incoming wave packet \eqref{eq:phi_in} and linear modulation \eqref{eq:potential}, can be used to compute the transmitted charge from our general expression \eqref{eq:Q}.
Taking into account that  $\mathcal{V}(t) = V_0- \alpha L_g/( 2 v)+ \alpha t$ for $u(x)$ given by Eq.~\eqref{eq:potential} 
 [computed in the limit $l_g \ll L_g$ with $x_b=0$ and $x_g \to +\infty$ in Eq.~\eqref{eq:uconv}; corresponds to the effective gate edge position $\tilde{x}_g = -L_g/2$ in contrast to $\tilde{x}_g=0$ and $\mathcal{V}(t)=V(t) $ used in Eq.~\eqref{eq:W_lin}], the extrapolation of our analytic theory to the one-dimensional model defined by $\hat{H}_{\text{1D}}$ reads
\begin{align} \label{eq:Qapprox}
  Q/e = \frac{1}{\tilde{\sigma}_E \sqrt{2 \pi }}\int  e^{-[E -E_0+V_0-\alpha L_g/( 2 v)]^2/(2 \tilde{\sigma}_E^2)}   T(E)  \, d E   \, ,
\end{align}   
where $\tilde{\sigma}_E  = \sqrt{\sigma_E^2+ (\hbar^2 \alpha^2/([4 \sigma_E^2])}$. Equation \eqref{eq:Qapprox} includes 
the transmission probability $T(E)$ for the static potential $V_c \, u(x)$; we compute $T(E)$ numerically using the scattering matrix expression in terms of a numerically computed Green function as in Ref.~\cite{Aharony02PRL}.

Figures \ref{fig:results1D}(b)-(f) plot the probability of transmission $Q/e$ for the same initial wave packet 
($k_0 =k_c$ and $\sigma_E=10^{-3}$), but different sharpness ($l_g=2$ versus $l_g=6$) and modulation speed ($\alpha=0$, $5 \cdot 10^{-6}$, and
$\pm 5 \cdot 10^{-5}$) of the barrier, as function of the additional barrier height $V_0$.
We note rather accurate agreement  between the direct numerical integration (dots) and the approximation \eqref{eq:Qapprox} (continuous blue line), validating the main thesis of this Section.

Additionally, we plot the transmission $T(E)$ as function of energy (dashed line).
For small $\tilde{\sigma}_E$, see Figures \ref{fig:results1D}(b),(e), the transmitted charge follows $T(E)$ since 
Since \eqref{eq:Qapprox} is a convolution of $T(E)$ and a Gaussian function of width $\tilde{\sigma}_E$,
the transmitted charge follows $T(E)$. Note that this agreement is a direct demonstration of the relation $T(E_0,V_0) =T(E_0-V_0)$ 
for the potential shown in Figures \ref{fig:results1D}(a) because $T$ is computed by varying $E_0$, and $Q$ is computed by varying $V_0$. 

 The sharper shape of $T(E)$ in Figs.\ \ref{fig:results1D}(b)-(d) corresponds to a steeper potential edge ($l_g=2$) compared to a wider step-like $T(E)$ for $l_g=6$ 
in Fig.~\ref{fig:results1D}(e),(f). Transmission maxima 
arise due to resonances at quasi-bound state energies $E_{0} = \pi^2 \hbar^2 n^2/(2 m^{\ast}L_g^2)$ where $n=1,2 \ldots$ is the number of halfwavelengths matching the barrier length $L_g$. We note that by increasing $L_g$ and adjusting the rounding parameter $l_g(L_g) \ll L_g$ accordignly, one can  achieve a smooth transition from $T=0$ to $T=1$ over an arbitrarily narrow energy range, suitable for modified Wigner function filtering as assumed in the examples of Section \ref{sec:results} and discussed in the context of the tomography technique below in Section~\ref{sec:experiment}.

\section{Connection to experiment\label{sec:experiment}}
In this section we discuss the relevance of our results to the recently proposed and implemented tomography protocol for solitary electrons~\cite{fletcher_2018}. 

\subsection{Scattering by a linearly modulated barrier as a Radon transform}
As mentioned in the introduction, the tomography experiment provides a map of the charge $Q(V_0,\alpha)$ transmitted through barrier that is controled by a gate voltage $V(t)=\alpha t + V_0$ with a linear time dependence as a function of the energy offset $V_0$ and the energy-time slope $\alpha=
\alpha_0 \, \tan \theta$, where $\alpha_0$ is a scale factor adapted to the characteristic time and energy scales of the experiment. The quantity
\begin{gather}
e^{-1} \frac{\partial Q(V_0,\theta)}{\partial V_0}
 = \int dt \, dE \frac{\partial T(E-V_0-\alpha_0 \, t \tan\theta)}{\partial V_0} \rho_{\rm in}(E,t) \nonumber \\
 = \int  dt \int d E \,  \frac{d T(E)}{d E} \, \rho_{\rm in}(V_0+\alpha_0 \, t\tan\theta-E,t) \label{eq:radon}
\end{gather}
represents the Radon transform \cite{radon} of the incoming Wigner distribution $\rho_{\rm in}(E,t)$ convolved with the energy derivative of the transmission coefficient $d T/ d E$. In Eq.~\eqref{eq:radon} we have used the fact that, according to Eq.~\eqref{eq:linearRes}, for a linear ramp the modified Wigner function $\tilde{\rho}(E,t)$ can be expressed in terms of the incoming Wigner function $\rho_{\rm in}(E,t)$ simply by adding $V(t)$ to the energy. 
The tomography protocol of Ref.~\cite{fletcher_2018} consists of measuring  \eqref{eq:radon} for a sufficiently wide range of $V_0$ and $\theta$ to enable numerical computation of the 
the inverse Radon transform using the standard filtered back-projection algorithm~\cite{radon}. A sufficiently sharp $T(E)$, such that the derivative $dT/dE$ can be approximated as a delta function, ensures that the inversion accurately represents the unknown $\rho_{\rm in}(E,t)$.

\begin{figure}
\center
\includegraphics[width=0.4\textwidth]{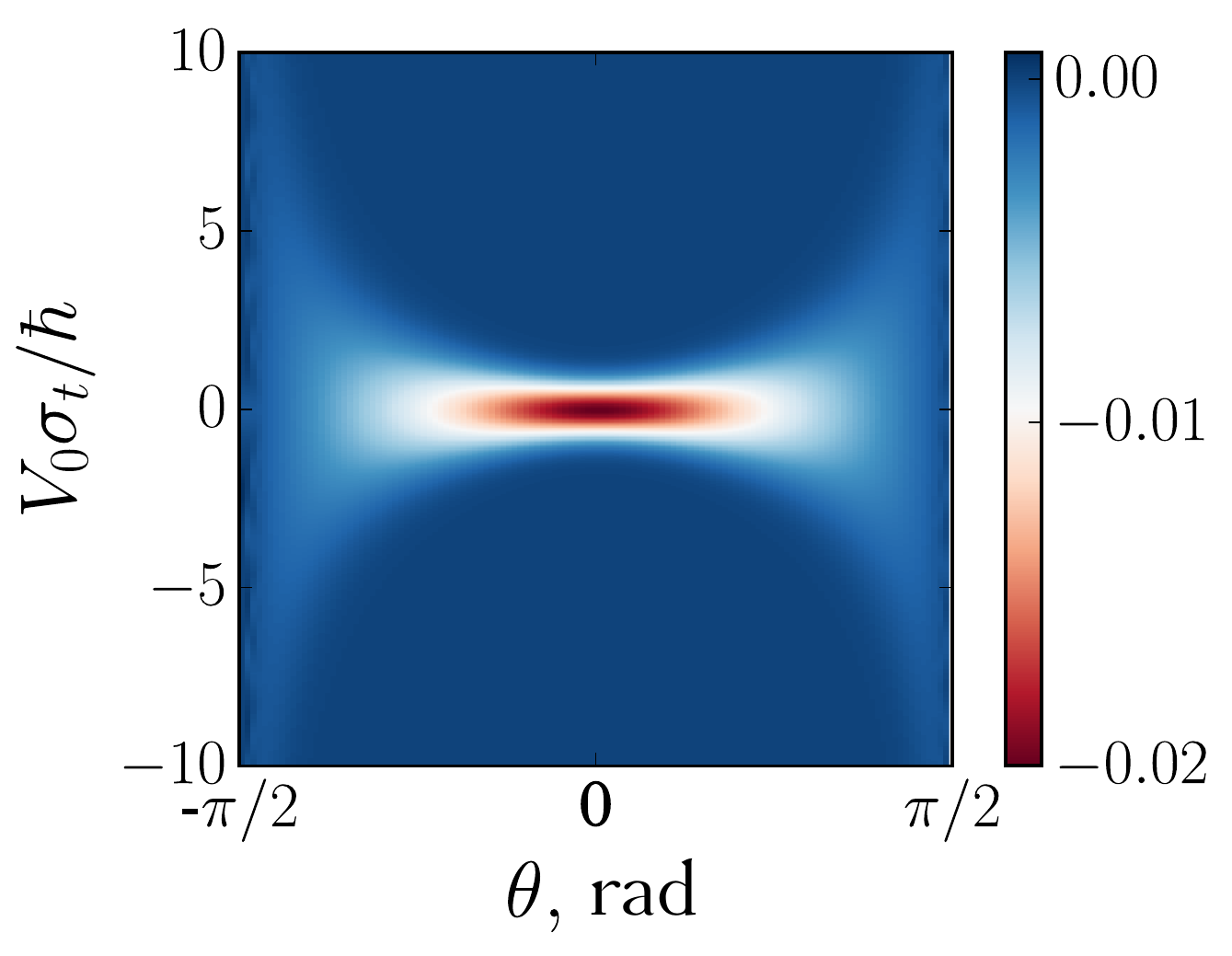}
\caption{\coloronline{(color online)} The Radon transform, defined in Eq.\ \eqref{eq:radon}, of the Wigner distribution of a Gaussian wave packet $\rho_{\rm in}(E,t)$  described by Eq.\ \eqref{eq:rho_in_gauss} with a step-function $T(E)$ ($\delta \to 0$). We have set $\alpha_0=\hbar/\sigma_t^2$.}
\label{fig:sinogram}
\end{figure}

To illustrate the Radon transform, in Fig.~\ref{fig:sinogram} we have shown the quantity $\partial Q(V_0,\theta)/\partial V_0$ calculated for the incoming Wigner distribution $\rho_{\rm in}(E,t)$ of a Gaussian wave packet, exactly like the one depicted in the top left panel of Fig.~\ref{fig:W_lin}.  The Radon transform is the line integral of $\rho_{\rm in}(E,t)$ along the straight line $E=V_0+ \alpha_0 t \tan\theta$. The non-zero values in Fig.~\ref{fig:sinogram} are concentrated around $\theta=0$. A correlation between energy and time \cite{slava_peter} that has a slope $\beta$ in the time-energy plane would introduce a shift by $\arctan (\beta/\alpha_0)$ along the $\theta$ axis \cite{fletcher_2018}. Notice that the width of the non-zero regions along the $V_0$ axis at $\theta=0$ corresponds to the quantum-limited energy width $\hbar/(2 \sigma_t)$ of the Gaussian distribution, as $\delta \to 0$ in this example.

\subsection{Resolution limits}
A finite sharpness $\delta$ of the transmission function $T(E)$ in Eq.~\eqref{eq:radon} can be seen as a finite resolution of the measurement apparatus that defines the energy level against which $\rho_{\text{in}}(E,t)$ is inferred by the tomography protocol. For $d T(E)/d E \ge 0$, the convolution of the incoming  Wigner function $\rho_{\rm in}(E,t)$ with $\partial T/\partial E$ in \eqref{eq:radon} can be interpreted as the Wigner representation
\begin{equation} \label{eq:effectiveRho}
\rho_{\text{eff}} (E,t) = (\pi \hbar)^{-1} \int  e^{i 2 E \, t'  / \hbar} \bra{t+t'} \hat{\rho} \ket{ t- t'} d t
\end{equation}
of an effective density matrix
\begin{align}
    \av{t | \hat{\rho} | t'} & =\int d E \frac{d T(E)}{d E} \varphi_E(t) \varphi^{\ast}_E(t') \\
    & = 
    \varphi_{\text{in}}(t) \, \Pi(t-t') \, \varphi^{\ast}_{\text{in}}(t') \, , \label{eq:rhoforms}
\end{align}
where
 $\varphi_E(t)=\varphi_{\text{in}}(t) e^{-i E t/\hbar}$ is the incoming state $\varphi_{\text{in}}(t)$ shifted in energy by $E$, and $\Pi(t)= \int  e^{- i E t/\hbar}  dT(E)$.
In the limit of $\delta \to 0$, $d T/d E$ becomes a delta-function, $\Pi(t-t')$ becomes $1$, and $\rho_{\text{eff}} (E,t)$ in  
Eq.~\eqref{eq:effectiveRho} reduces to the pure-state Wigner function \eqref{eq:W_def}. For finite $\delta$ though,  the characteristic temporal width $\hbar / \delta$ of the Fourier transform $\Pi(t-t')$ in the time-domain representation [Eq.~\eqref{eq:rhoforms}]  sets the upper limit on the coherence time of the wave-packet that can be resolved by the tomographic reconstruction.

For $\delta>0$, electron partitioning at the barrier introduces shot noise that reduces the purity of the reconstructed effective mixed state $\hat{\rho}$. 
We illustrate this by computing the convolution of the uncorrelated Gaussian wavepacket \eqref{eq:phi_in} and a Gaussian $dT/dE$ from Eq.~\eqref{eq:T},  
\begin{align}
& \rho_{\text{eff}}(E,t)=  \label{eq:rhoeff} \\
& \frac{1}{2 \pi\sqrt{(\hbar/2)^2 + (\sigma_t \, \delta)^2}}
\exp \left \{ 
-\frac{t^2}{2\sigma_t^2} 
- \frac{1}{2} \frac{(E-E_0)^2}{[\hbar/(2 \sigma_t)]^2+\delta^2} 
\right \} . \nonumber
\end{align}
Quantum purity, defined as $\gamma = \tr \rho^2$, is straightforward to compute in the  Wigner representation, 
$\gamma=  h \iint \rho_{\text{eff}}(E,t) \, dE \, dt$ \cite{fletcher_2018}. Using 
\eqref{eq:rhoeff} we obtain  the effective purity of tomographic reconstruction by a finite-width barrier for a quantum-limited Gaussian wave-packet,
\begin{align}
   \gamma=\frac{\hbar/2}{\sqrt{(\hbar/2)^2 + (\sigma_t \, \delta)^2}} \, .
\end{align}
We see that $\gamma$ in the case of an uncorrelated Gaussian wavepacket is equal to the ratio between the ideal Heisenberg uncertainty product $(\hbar/2)$ and the product of time and energy widths in the effective distribution \eqref{eq:rhoeff} which is broadened by a finite resolution $\delta$ of the energy detector.
Note that in general, the uncertainty product alone is insufficient to distinguish the incoherent broadening (e.g.
$\delta$-dependent terms in Eq.~\eqref{eq:rhoeff}] from coherent broadening (e.g. the energy-time correlation created in the source~\cite{slava_peter,fletcher_2018}). Hence a full tomography technique is essential for evaluating the quality a  source with respect to the quantum limit of localizing a single particle in energy and time.

The dependence of $\delta$ on the parameters of the constriction depends on the details of the scattering interaction $\hat H_{\rm bs}$ and goes beyong the scope of this work. Nevertheless, we can use the simple sketch in Fig.~\ref{fig:setup_exp2} for an order of magnitude estimate.
For a single-channel model with the backscattering amplitude distributed spatially over a length $x \in [-x_b \ldots +x_b]$, the temporal width of $\tau(t)$ (and hence $\Pi(t)$)  is limited by $2 \, x_b/v$. This gives an estimate
 \begin{align} \label{eq:lengthlimit}
 \delta \gtrsim \frac{\hbar v}{x_b}
\end{align}
for the achievable energy resolution of the tomography technique within a specific model of the scattering barrier depicted schematically in Fig.~\ref{fig:setup_exp2}.
This bound can also be interpreted as $x_b \, \Delta p  \gtrsim \hbar$ where  $\Delta p = \delta /v$ is the Fourier (diffraction) limit on the momentum resolution achievable over the characteristic length $x_b$ of the backscattering region for a particle propagating with velocity $v$.

One can expect from Eq.~\eqref{eq:lengthlimit} that the effective length of the spatially uniform part of the scattering region is an important parameter for designing a barrier suitable for high-resolution tomography. On the other hand, as we have seen in Sec.~\ref{sec:SpecialCases}, the length and the shape
of the gate edge region from $-x_g$ to $-x_b$ is not important as long as the linear-in-time modulation condition \eqref{eq:linear} is fulfilled and the velocity dispersion can be ignored.

\subsection{Measurement of the modified Wigner function}
As mentioned in Sec.\ \ref{sec:derivation}, the modified Wigner function $\tilde{\rho}(E,t)$ is not a physical observable and does not represent the Wigner function of the wave packet at any stage. However, it can be experimentally obtained as the outcome of tomographic reconstruction. To see this, we consider a gate voltage $V(t)$ that is the sum of a linear part $V_0+\alpha \, t$, as in the experiment of Ref. \cite{fletcher_2018}, and an additional perturbation $V'(t)$. This scenario is discussed theoretically at the end of Sec.\ \ref{sec:SpecialCases}, see Eqs.\ \eqref{eq:V_prime} and \eqref{eq:rho_prime}. Analogously to Eq.\ \eqref{eq:radon}, the modified Wigner function corresponding to the perturbation $V'(t)$ only, can be expressed as 
\begin{align}
\frac{\partial Q(V_0,\theta)}{\partial V_0}
& = e \int dt \, \tilde{\rho}'(V_0+\alpha_0 \, t \tan\theta,t),
\end{align}
where we have assumed ideal energy resolution, inserted $\mathcal{V}(t)=V_0+\alpha_0 \, t \tan \theta$, and absorbed the correction to $V(t)$ due to a possibly non-sharp edge of the gate into $V_0$. This result means that, if we perform the inverse Radon transform of ${\partial Q(V_0,\theta)}/{\partial V_0}$, the outcome would be the modified Wigner distribution calculated for the non-linear part $V'(t)$ of the potential $V(t)$ only. For example, the function $\tilde{\rho}(E,t)$ of Eq.\ \eqref{eq:rho_step} can be measured using the linear tomographic reconstruction protocol of Ref. \cite{fletcher_2018} while applying a gate voltage that is the sum of Eq.\ \eqref{eq:V_lin} and Eq.\ \eqref{eq:V_step}, i.e.,
\begin{align}
V(t) = \left\{ \begin{array}{ll} V_0+\alpha t + V_{\rm i} & \mbox{for $t < t_0$}, \\ 
  V_0+\alpha t+V_{\rm f} & \mbox{for $t > t_0$}. \end{array} \right.
\end{align}
By introducing such a sharp voltage ``kick'' on top of a linear time dependence of the gate potential, it  would be possible to gather direct evidence of coherence of the electron source, since negative values in the measurement of $\tilde{\rho}'(E,t)$ arise from interference between phase-coherent parts of the incoming wave packet. This is relevant for the investigation of the properties of single-electron sources that emit solitary electrons high above the Fermi energy, where coherence is key to potential interferometric applications~\cite{bauerle_2018}.

Our results also provide a way to describe a different kind of problem: sampling the potential $V(t)$ with single electrons whose incoming distribution is well known~\cite{johnson_2017}. In this case, the deviations in the measurement of $\rho_{\rm in}(E,t)$ from the actual incoming distribution may allow one to measure the wave-form $V(t)$ beyond the classical resolution limit. In this context, Eq.~\eqref{eq:uconv} can elucidate the influence of the potential shape on the bandwidth of such a ``quantum oscilloscopic'' measurement.

\section{Conclusions}\label{sec:Conclusion}
Inspired by recent experiments~\cite{kataoka_pss,fletcher_2018}, we have constructed a fully quantum-mechanical description of a dynamical scattering problem of electron wave packets in a one-dimensional chiral channel passing through a constriction subject to a time-dependent gate potential $V(t)$. We have shown that the expression for the transmitted charge in this system is analogous to the corresponding classical expression, with a modified Wigner distribution function $\tilde \rho$ being the quantum analog of the classical probability distribution. In particular, if the gate voltage time dependence is linear then the modified Wigner distribution is obtained from the Wigner function of the incoming electron by a time-dependent shift of the energy $E$. In this case the full quantum-mechanical theory agrees with the classical-limit estimate (\ref{eq:Q_semicl}) justifying the quantum tomography proposal put forward in 
Ref.~\onlinecite{fletcher_2018}.

A finite energy resolution $\delta$ of the scattering barrier adds shot noise to the tomographically reconstructed Wigner distribution,  limiting the maximal coherence time that can be probed by $\hbar /\delta$. Small $\delta$ requires large spatial extent $x_b$ of the uniformly modulated part of the constriction, $ \delta \geq  \hbar v/ x_b$. We have shown that neither the length of the wave-packet nor the shape of the accelerating edge of the gate potential perturbs the linear tomography protocol as long as the modulation of the scattering region remains spatially uniform.

Our fully quantum-mechanical theory provides the tools to analyze experiments with a general time dependence of the gate voltage, relevant for wave-packets that are stretched in the time domain. 
The results presented here can be used to describe existing and propose new quantum measurement protocols. They can also serve to analyze a reverse problem: determining the gate potential $V(t)$ when the incoming electron distribution is known. 
We have shown that additional modification of the Wigner function  by non-linear gate modulation can in principle be revealed by quantum tomography within the same device.  This opens a possibility to probe coherence of single-electron sources   with explicit signatures of quantum interference (e.g., negativity of the modified Wigner function) which does not require multiple-path layouts.

\subsection*{Acknowledgements}
We gratefully acknowledge support from the German Research Foundation (DFG) in the framework of the Collaborative Research Center 658, from the Latvian Council of Science (grant No.~lzp-2018/1-0173), and from University of Latvia (grant no.~AAP2016/B031). This work was supported in part by the Joint Research Project SEQUOIA (17FUN04). This project has received funding from the European Metrology Programme for Innovation and Research (EMPIR) co-financed by the Participating States and from the European Union's Horizon 2020 research and innovation programme. We thank J.~D.~Fletcher, M.~Kataoka, E.~Khalaf, and C.~Fr\"{a}{\ss}dorf for valuable discussions and C.~Kl\"{o}ckner for help with numerical integration.

\bibliographystyle{apsrev4-1}

\begin{thebibliography}{48}%
\makeatletter
\providecommand \@ifxundefined [1]{%
 \@ifx{#1\undefined}
}%
\providecommand \@ifnum [1]{%
 \ifnum #1\expandafter \@firstoftwo
 \else \expandafter \@secondoftwo
 \fi
}%
\providecommand \@ifx [1]{%
 \ifx #1\expandafter \@firstoftwo
 \else \expandafter \@secondoftwo
 \fi
}%
\providecommand \natexlab [1]{#1}%
\providecommand \enquote  [1]{``#1''}%
\providecommand \bibnamefont  [1]{#1}%
\providecommand \bibfnamefont [1]{#1}%
\providecommand \citenamefont [1]{#1}%
\providecommand \href@noop [0]{\@secondoftwo}%
\providecommand \href [0]{\begingroup \@sanitize@url \@href}%
\providecommand \@href[1]{\@@startlink{#1}\@@href}%
\providecommand \@@href[1]{\endgroup#1\@@endlink}%
\providecommand \@sanitize@url [0]{\catcode `\\12\catcode `\$12\catcode
  `\&12\catcode `\#12\catcode `\^12\catcode `\_12\catcode `\%12\relax}%
\providecommand \@@startlink[1]{}%
\providecommand \@@endlink[0]{}%
\providecommand \url  [0]{\begingroup\@sanitize@url \@url }%
\providecommand \@url [1]{\endgroup\@href {#1}{\urlprefix }}%
\providecommand \urlprefix  [0]{URL }%
\providecommand \Eprint [0]{\href }%
\providecommand \doibase [0]{http://dx.doi.org/}%
\providecommand \selectlanguage [0]{\@gobble}%
\providecommand \bibinfo  [0]{\@secondoftwo}%
\providecommand \bibfield  [0]{\@secondoftwo}%
\providecommand \translation [1]{[#1]}%
\providecommand \BibitemOpen [0]{}%
\providecommand \bibitemStop [0]{}%
\providecommand \bibitemNoStop [0]{.\EOS\space}%
\providecommand \EOS [0]{\spacefactor3000\relax}%
\providecommand \BibitemShut  [1]{\csname bibitem#1\endcsname}%
\let\auto@bib@innerbib\@empty
\bibitem [{\citenamefont {Grenier}\ \emph
  {et~al.}(2011{\natexlab{a}})\citenamefont {Grenier}, \citenamefont
  {Herv\'{e}}, \citenamefont {F\`{e}ve},\ and\ \citenamefont
  {Degiovanni}}]{degiovanni_rev_2011}%
  \BibitemOpen
  \bibfield  {author} {\bibinfo {author} {\bibfnamefont {C.}~\bibnamefont
  {Grenier}}, \bibinfo {author} {\bibfnamefont {R.}~\bibnamefont {Herv\'{e}}},
  \bibinfo {author} {\bibfnamefont {G.}~\bibnamefont {F\`{e}ve}}, \ and\
  \bibinfo {author} {\bibfnamefont {P.}~\bibnamefont {Degiovanni}},\ }\href
  {\doibase 10.1142/S0217984911026772} {\bibfield  {journal} {\bibinfo
  {journal} {Mod. Phys. Lett. B}\ }\textbf {\bibinfo {volume} {25}},\ \bibinfo
  {pages} {1053} (\bibinfo {year} {2011}{\natexlab{a}})}\BibitemShut {NoStop}%
\bibitem [{\citenamefont {Bocquillon}\ \emph
  {et~al.}(2014{\natexlab{a}})\citenamefont {Bocquillon}, \citenamefont
  {Freulon}, \citenamefont {Parmentier}, \citenamefont {Berroir}, \citenamefont
  {Pla\c{c}ais}, \citenamefont {Wahl}, \citenamefont {Rech}, \citenamefont
  {Jonckheere}, \citenamefont {Martin}, \citenamefont {Grenier}, \citenamefont
  {Ferraro}, \citenamefont {Degiovanni},\ and\ \citenamefont
  {F\`eve}}]{bocquillon_rev}%
  \BibitemOpen
  \bibfield  {author} {\bibinfo {author} {\bibfnamefont {E.}~\bibnamefont
  {Bocquillon}}, \bibinfo {author} {\bibfnamefont {V.}~\bibnamefont {Freulon}},
  \bibinfo {author} {\bibfnamefont {F.~D.}\ \bibnamefont {Parmentier}},
  \bibinfo {author} {\bibfnamefont {J.-M.}\ \bibnamefont {Berroir}}, \bibinfo
  {author} {\bibfnamefont {B.}~\bibnamefont {Pla\c{c}ais}}, \bibinfo {author}
  {\bibfnamefont {C.}~\bibnamefont {Wahl}}, \bibinfo {author} {\bibfnamefont
  {J.}~\bibnamefont {Rech}}, \bibinfo {author} {\bibfnamefont {T.}~\bibnamefont
  {Jonckheere}}, \bibinfo {author} {\bibfnamefont {T.}~\bibnamefont {Martin}},
  \bibinfo {author} {\bibfnamefont {C.}~\bibnamefont {Grenier}}, \bibinfo
  {author} {\bibfnamefont {D.}~\bibnamefont {Ferraro}}, \bibinfo {author}
  {\bibfnamefont {P.}~\bibnamefont {Degiovanni}}, \ and\ \bibinfo {author}
  {\bibfnamefont {G.}~\bibnamefont {F\`eve}},\ }\href {\doibase
  10.1002/andp.201300181} {\bibfield  {journal} {\bibinfo  {journal} {Annalen
  der Physik}\ }\textbf {\bibinfo {volume} {526}},\ \bibinfo {pages} {1}
  (\bibinfo {year} {2014}{\natexlab{a}})}\BibitemShut {NoStop}%
\bibitem [{\citenamefont {B{\"a}uerle}\ \emph {et~al.}(2018)\citenamefont
  {B{\"a}uerle}, \citenamefont {Glattli}, \citenamefont {Meunier},
  \citenamefont {Portier}, \citenamefont {Roche}, \citenamefont {Roulleau},
  \citenamefont {Takada},\ and\ \citenamefont {Waintal}}]{bauerle_2018}%
  \BibitemOpen
  \bibfield  {author} {\bibinfo {author} {\bibfnamefont {C.}~\bibnamefont
  {B{\"a}uerle}}, \bibinfo {author} {\bibfnamefont {C.}~\bibnamefont
  {Glattli}}, \bibinfo {author} {\bibfnamefont {T.}~\bibnamefont {Meunier}},
  \bibinfo {author} {\bibfnamefont {F.}~\bibnamefont {Portier}}, \bibinfo
  {author} {\bibfnamefont {P.}~\bibnamefont {Roche}}, \bibinfo {author}
  {\bibfnamefont {P.}~\bibnamefont {Roulleau}}, \bibinfo {author}
  {\bibfnamefont {S.}~\bibnamefont {Takada}}, \ and\ \bibinfo {author}
  {\bibfnamefont {X.}~\bibnamefont {Waintal}},\ }\href {\doibase
  10.1088/1361-6633/aaa98a} {\bibfield  {journal} {\bibinfo  {journal} {Reports
  on Progress in Physics}\ }\textbf {\bibinfo {volume} {81}},\ \bibinfo {pages}
  {056503} (\bibinfo {year} {2018})}\BibitemShut {NoStop}%
\bibitem [{\citenamefont {Marguerite}\ \emph {et~al.}(2017)\citenamefont
  {Marguerite}, \citenamefont {Bocquillon}, \citenamefont {Berroir},
  \citenamefont {Pla\c{c}ais}, \citenamefont {Cavanna}, \citenamefont {Jin},
  \citenamefont {Degiovanni},\ and\ \citenamefont {F\`eve}}]{feve_bocq_rev}%
  \BibitemOpen
  \bibfield  {author} {\bibinfo {author} {\bibfnamefont {A.}~\bibnamefont
  {Marguerite}}, \bibinfo {author} {\bibfnamefont {E.}~\bibnamefont
  {Bocquillon}}, \bibinfo {author} {\bibfnamefont {J.-M.}\ \bibnamefont
  {Berroir}}, \bibinfo {author} {\bibfnamefont {B.}~\bibnamefont
  {Pla\c{c}ais}}, \bibinfo {author} {\bibfnamefont {A.}~\bibnamefont
  {Cavanna}}, \bibinfo {author} {\bibfnamefont {Y.}~\bibnamefont {Jin}},
  \bibinfo {author} {\bibfnamefont {P.}~\bibnamefont {Degiovanni}}, \ and\
  \bibinfo {author} {\bibfnamefont {G.}~\bibnamefont {F\`eve}},\ }\href
  {\doibase 10.1002/pssb.201600618} {\bibfield  {journal} {\bibinfo  {journal}
  {Physica Status Solidi B}\ }\textbf {\bibinfo {volume} {254}},\ \bibinfo
  {pages} {1600618} (\bibinfo {year} {2017})}\BibitemShut {NoStop}%
\bibitem [{\citenamefont {Roussel}\ \emph {et~al.}(2017)\citenamefont
  {Roussel}, \citenamefont {Cabart}, \citenamefont {F\`eve}, \citenamefont
  {Thibierge},\ and\ \citenamefont {Degiovanni}}]{feve_rev}%
  \BibitemOpen
  \bibfield  {author} {\bibinfo {author} {\bibfnamefont {B.}~\bibnamefont
  {Roussel}}, \bibinfo {author} {\bibfnamefont {C.}~\bibnamefont {Cabart}},
  \bibinfo {author} {\bibfnamefont {G.}~\bibnamefont {F\`eve}}, \bibinfo
  {author} {\bibfnamefont {E.}~\bibnamefont {Thibierge}}, \ and\ \bibinfo
  {author} {\bibfnamefont {P.}~\bibnamefont {Degiovanni}},\ }\href {\doibase
  10.1002/pssb.201600621} {\bibfield  {journal} {\bibinfo  {journal} {Physica
  Status Solidi B}\ }\textbf {\bibinfo {volume} {254}},\ \bibinfo {pages}
  {1600621} (\bibinfo {year} {2017})}\BibitemShut {NoStop}%
\bibitem [{\citenamefont {{Demkowicz-Dobrza{\'n}ski}}\ \emph
  {et~al.}(2012)\citenamefont {{Demkowicz-Dobrza{\'n}ski}}, \citenamefont
  {{Ko{\l}ody{\'n}ski}},\ and\ \citenamefont {{Gu{\c t}{\u
  a}}}}]{demkowicz_2012}%
  \BibitemOpen
  \bibfield  {author} {\bibinfo {author} {\bibfnamefont {R.}~\bibnamefont
  {{Demkowicz-Dobrza{\'n}ski}}}, \bibinfo {author} {\bibfnamefont
  {J.}~\bibnamefont {{Ko{\l}ody{\'n}ski}}}, \ and\ \bibinfo {author}
  {\bibfnamefont {M.}~\bibnamefont {{Gu{\c t}{\u a}}}},\ }\href {\doibase
  10.1038/ncomms2067} {\bibfield  {journal} {\bibinfo  {journal} {Nature
  Communications}\ }\textbf {\bibinfo {volume} {3}},\ \bibinfo {eid} {1063}
  (\bibinfo {year} {2012})}\BibitemShut {NoStop}%
\bibitem [{\citenamefont {Pothier}\ \emph {et~al.}(1992)\citenamefont
  {Pothier}, \citenamefont {Lafarge}, \citenamefont {Urbina}, \citenamefont
  {Esteve},\ and\ \citenamefont {Devoret}}]{pothier_1992}%
  \BibitemOpen
  \bibfield  {author} {\bibinfo {author} {\bibfnamefont {H.}~\bibnamefont
  {Pothier}}, \bibinfo {author} {\bibfnamefont {P.}~\bibnamefont {Lafarge}},
  \bibinfo {author} {\bibfnamefont {C.}~\bibnamefont {Urbina}}, \bibinfo
  {author} {\bibfnamefont {D.}~\bibnamefont {Esteve}}, \ and\ \bibinfo {author}
  {\bibfnamefont {M.~H.}\ \bibnamefont {Devoret}},\ }\href {\doibase
  10.1209/0295-5075/17/3/011} {\bibfield  {journal} {\bibinfo  {journal}
  {Europhys. Lett.}\ }\textbf {\bibinfo {volume} {17}},\ \bibinfo {pages} {249}
  (\bibinfo {year} {1992})}\BibitemShut {NoStop}%
\bibitem [{\citenamefont {Switkes}\ \emph {et~al.}(1999)\citenamefont
  {Switkes}, \citenamefont {Marcus}, \citenamefont {Campman},\ and\
  \citenamefont {Gossard}}]{marcus_1999}%
  \BibitemOpen
  \bibfield  {author} {\bibinfo {author} {\bibfnamefont {M.}~\bibnamefont
  {Switkes}}, \bibinfo {author} {\bibfnamefont {C.~M.}\ \bibnamefont {Marcus}},
  \bibinfo {author} {\bibfnamefont {K.}~\bibnamefont {Campman}}, \ and\
  \bibinfo {author} {\bibfnamefont {A.~C.}\ \bibnamefont {Gossard}},\ }\href
  {\doibase 10.1126/science.283.5409.1905} {\bibfield  {journal} {\bibinfo
  {journal} {Science}\ }\textbf {\bibinfo {volume} {283}},\ \bibinfo {pages}
  {1905} (\bibinfo {year} {1999})}\BibitemShut {NoStop}%
\bibitem [{\citenamefont {{F{\`e}ve}}\ \emph {et~al.}(2007)\citenamefont
  {{F{\`e}ve}}, \citenamefont {{Mah{\'e}}}, \citenamefont {{Berroir}},
  \citenamefont {{Kontos}}, \citenamefont {{Pla{\c c}ais}}, \citenamefont
  {{Glattli}}, \citenamefont {{Cavanna}}, \citenamefont {{Etienne}},\ and\
  \citenamefont {{Jin}}}]{feve_2007}%
  \BibitemOpen
  \bibfield  {author} {\bibinfo {author} {\bibfnamefont {G.}~\bibnamefont
  {{F{\`e}ve}}}, \bibinfo {author} {\bibfnamefont {A.}~\bibnamefont
  {{Mah{\'e}}}}, \bibinfo {author} {\bibfnamefont {J.-M.}\ \bibnamefont
  {{Berroir}}}, \bibinfo {author} {\bibfnamefont {T.}~\bibnamefont {{Kontos}}},
  \bibinfo {author} {\bibfnamefont {B.}~\bibnamefont {{Pla{\c c}ais}}},
  \bibinfo {author} {\bibfnamefont {D.~C.}\ \bibnamefont {{Glattli}}}, \bibinfo
  {author} {\bibfnamefont {A.}~\bibnamefont {{Cavanna}}}, \bibinfo {author}
  {\bibfnamefont {B.}~\bibnamefont {{Etienne}}}, \ and\ \bibinfo {author}
  {\bibfnamefont {Y.}~\bibnamefont {{Jin}}},\ }\href {\doibase
  10.1126/science.1141243} {\bibfield  {journal} {\bibinfo  {journal}
  {Science}\ }\textbf {\bibinfo {volume} {316}},\ \bibinfo {pages} {1169}
  (\bibinfo {year} {2007})}\BibitemShut {NoStop}%
\bibitem [{\citenamefont {{Blumenthal}}\ \emph {et~al.}(2007)\citenamefont
  {{Blumenthal}}, \citenamefont {{Kaestner}}, \citenamefont {{Li}},
  \citenamefont {{Giblin}}, \citenamefont {{Janssen}}, \citenamefont
  {{Pepper}}, \citenamefont {{Anderson}}, \citenamefont {{Jones}},\ and\
  \citenamefont {{Ritchie}}}]{kaestner_2007}%
  \BibitemOpen
  \bibfield  {author} {\bibinfo {author} {\bibfnamefont {M.~D.}\ \bibnamefont
  {{Blumenthal}}}, \bibinfo {author} {\bibfnamefont {B.}~\bibnamefont
  {{Kaestner}}}, \bibinfo {author} {\bibfnamefont {L.}~\bibnamefont {{Li}}},
  \bibinfo {author} {\bibfnamefont {S.}~\bibnamefont {{Giblin}}}, \bibinfo
  {author} {\bibfnamefont {T.~J.~B.~M.}\ \bibnamefont {{Janssen}}}, \bibinfo
  {author} {\bibfnamefont {M.}~\bibnamefont {{Pepper}}}, \bibinfo {author}
  {\bibfnamefont {D.}~\bibnamefont {{Anderson}}}, \bibinfo {author}
  {\bibfnamefont {G.}~\bibnamefont {{Jones}}}, \ and\ \bibinfo {author}
  {\bibfnamefont {D.~A.}\ \bibnamefont {{Ritchie}}},\ }\href {\doibase
  10.1038/nphys582} {\bibfield  {journal} {\bibinfo  {journal} {Nature
  Physics}\ }\textbf {\bibinfo {volume} {3}},\ \bibinfo {pages} {343} (\bibinfo
  {year} {2007})}\BibitemShut {NoStop}%
\bibitem [{\citenamefont {Fujiwara}\ \emph {et~al.}(2008)\citenamefont
  {Fujiwara}, \citenamefont {Nishiguchi},\ and\ \citenamefont
  {Ono}}]{fujiwara_2008}%
  \BibitemOpen
  \bibfield  {author} {\bibinfo {author} {\bibfnamefont {A.}~\bibnamefont
  {Fujiwara}}, \bibinfo {author} {\bibfnamefont {K.}~\bibnamefont
  {Nishiguchi}}, \ and\ \bibinfo {author} {\bibfnamefont {Y.}~\bibnamefont
  {Ono}},\ }\href {\doibase 10.1063/1.2837544} {\bibfield  {journal} {\bibinfo
  {journal} {Applied Physics Letters}\ }\textbf {\bibinfo {volume} {92}},\
  \bibinfo {pages} {042102} (\bibinfo {year} {2008})}\BibitemShut {NoStop}%
\bibitem [{\citenamefont {{Pekola}}\ \emph {et~al.}(2008)\citenamefont
  {{Pekola}}, \citenamefont {{Vartiainen}}, \citenamefont {{M{\"o}tt{\"o}nen}},
  \citenamefont {{Saira}}, \citenamefont {{Meschke}},\ and\ \citenamefont
  {{Averin}}}]{pekola_2008}%
  \BibitemOpen
  \bibfield  {author} {\bibinfo {author} {\bibfnamefont {J.~P.}\ \bibnamefont
  {{Pekola}}}, \bibinfo {author} {\bibfnamefont {J.~J.}\ \bibnamefont
  {{Vartiainen}}}, \bibinfo {author} {\bibfnamefont {M.}~\bibnamefont
  {{M{\"o}tt{\"o}nen}}}, \bibinfo {author} {\bibfnamefont {O.-P.}\ \bibnamefont
  {{Saira}}}, \bibinfo {author} {\bibfnamefont {M.}~\bibnamefont {{Meschke}}},
  \ and\ \bibinfo {author} {\bibfnamefont {D.~V.}\ \bibnamefont {{Averin}}},\
  }\href {\doibase 10.1038/nphys808} {\bibfield  {journal} {\bibinfo  {journal}
  {Nature Physics}\ }\textbf {\bibinfo {volume} {4}},\ \bibinfo {pages} {120}
  (\bibinfo {year} {2008})}\BibitemShut {NoStop}%
\bibitem [{\citenamefont {Kaestner}\ \emph {et~al.}(2008)\citenamefont
  {Kaestner}, \citenamefont {Kashcheyevs}, \citenamefont {Amakawa},
  \citenamefont {Blumenthal}, \citenamefont {Li}, \citenamefont {Janssen},
  \citenamefont {Hein}, \citenamefont {Pierz}, \citenamefont {Weimann},
  \citenamefont {Siegner},\ and\ \citenamefont {Schumacher}}]{kaestner_2008}%
  \BibitemOpen
  \bibfield  {author} {\bibinfo {author} {\bibfnamefont {B.}~\bibnamefont
  {Kaestner}}, \bibinfo {author} {\bibfnamefont {V.}~\bibnamefont
  {Kashcheyevs}}, \bibinfo {author} {\bibfnamefont {S.}~\bibnamefont
  {Amakawa}}, \bibinfo {author} {\bibfnamefont {M.~D.}\ \bibnamefont
  {Blumenthal}}, \bibinfo {author} {\bibfnamefont {L.}~\bibnamefont {Li}},
  \bibinfo {author} {\bibfnamefont {T.~J. B.~M.}\ \bibnamefont {Janssen}},
  \bibinfo {author} {\bibfnamefont {G.}~\bibnamefont {Hein}}, \bibinfo {author}
  {\bibfnamefont {K.}~\bibnamefont {Pierz}}, \bibinfo {author} {\bibfnamefont
  {T.}~\bibnamefont {Weimann}}, \bibinfo {author} {\bibfnamefont
  {U.}~\bibnamefont {Siegner}}, \ and\ \bibinfo {author} {\bibfnamefont
  {H.~W.}\ \bibnamefont {Schumacher}},\ }\href {\doibase
  10.1103/PhysRevB.77.153301} {\bibfield  {journal} {\bibinfo  {journal} {Phys.
  Rev. B}\ }\textbf {\bibinfo {volume} {77}},\ \bibinfo {pages} {153301}
  (\bibinfo {year} {2008})}\BibitemShut {NoStop}%
\bibitem [{\citenamefont {Wright}\ \emph {et~al.}(2008)\citenamefont {Wright},
  \citenamefont {Blumenthal}, \citenamefont {Gumbs}, \citenamefont {Thorn},
  \citenamefont {Pepper}, \citenamefont {Janssen}, \citenamefont {Holmes},
  \citenamefont {Anderson}, \citenamefont {Jones}, \citenamefont {Nicoll},\
  and\ \citenamefont {Ritchie}}]{blumenthal_2008}%
  \BibitemOpen
  \bibfield  {author} {\bibinfo {author} {\bibfnamefont {S.~J.}\ \bibnamefont
  {Wright}}, \bibinfo {author} {\bibfnamefont {M.~D.}\ \bibnamefont
  {Blumenthal}}, \bibinfo {author} {\bibfnamefont {G.}~\bibnamefont {Gumbs}},
  \bibinfo {author} {\bibfnamefont {A.~L.}\ \bibnamefont {Thorn}}, \bibinfo
  {author} {\bibfnamefont {M.}~\bibnamefont {Pepper}}, \bibinfo {author}
  {\bibfnamefont {T.~J. B.~M.}\ \bibnamefont {Janssen}}, \bibinfo {author}
  {\bibfnamefont {S.~N.}\ \bibnamefont {Holmes}}, \bibinfo {author}
  {\bibfnamefont {D.}~\bibnamefont {Anderson}}, \bibinfo {author}
  {\bibfnamefont {G.~A.~C.}\ \bibnamefont {Jones}}, \bibinfo {author}
  {\bibfnamefont {C.~A.}\ \bibnamefont {Nicoll}}, \ and\ \bibinfo {author}
  {\bibfnamefont {D.~A.}\ \bibnamefont {Ritchie}},\ }\href {\doibase
  10.1103/PhysRevB.78.233311} {\bibfield  {journal} {\bibinfo  {journal} {Phys.
  Rev. B}\ }\textbf {\bibinfo {volume} {78}},\ \bibinfo {pages} {233311}
  (\bibinfo {year} {2008})}\BibitemShut {NoStop}%
\bibitem [{\citenamefont {Kaestner}\ \emph {et~al.}(2009)\citenamefont
  {Kaestner}, \citenamefont {Leicht}, \citenamefont {Kashcheyevs},
  \citenamefont {Pierz}, \citenamefont {Siegner},\ and\ \citenamefont
  {Schumacher}}]{kaestner_2009}%
  \BibitemOpen
  \bibfield  {author} {\bibinfo {author} {\bibfnamefont {B.}~\bibnamefont
  {Kaestner}}, \bibinfo {author} {\bibfnamefont {C.}~\bibnamefont {Leicht}},
  \bibinfo {author} {\bibfnamefont {V.}~\bibnamefont {Kashcheyevs}}, \bibinfo
  {author} {\bibfnamefont {K.}~\bibnamefont {Pierz}}, \bibinfo {author}
  {\bibfnamefont {U.}~\bibnamefont {Siegner}}, \ and\ \bibinfo {author}
  {\bibfnamefont {H.~W.}\ \bibnamefont {Schumacher}},\ }\href {\doibase
  10.1063/1.3063128} {\bibfield  {journal} {\bibinfo  {journal} {Applied
  Physics Letters}\ }\textbf {\bibinfo {volume} {94}},\ \bibinfo {pages}
  {012106} (\bibinfo {year} {2009})}\BibitemShut {NoStop}%
\bibitem [{\citenamefont {{Chorley}}\ \emph {et~al.}(2012)\citenamefont
  {{Chorley}}, \citenamefont {{Frake}}, \citenamefont {{Smith}}, \citenamefont
  {{Jones}},\ and\ \citenamefont {{Buitelaar}}}]{chorley_2012}%
  \BibitemOpen
  \bibfield  {author} {\bibinfo {author} {\bibfnamefont {S.~J.}\ \bibnamefont
  {{Chorley}}}, \bibinfo {author} {\bibfnamefont {J.}~\bibnamefont {{Frake}}},
  \bibinfo {author} {\bibfnamefont {C.~G.}\ \bibnamefont {{Smith}}}, \bibinfo
  {author} {\bibfnamefont {G.~A.~C.}\ \bibnamefont {{Jones}}}, \ and\ \bibinfo
  {author} {\bibfnamefont {M.~R.}\ \bibnamefont {{Buitelaar}}},\ }\href
  {\doibase 10.1063/1.3700967} {\bibfield  {journal} {\bibinfo  {journal}
  {Applied Physics Letters}\ }\textbf {\bibinfo {volume} {100}},\ \bibinfo
  {eid} {143104} (\bibinfo {year} {2012})}\BibitemShut {NoStop}%
\bibitem [{\citenamefont {{Connolly}}\ \emph {et~al.}(2013)\citenamefont
  {{Connolly}}, \citenamefont {{Chiu}}, \citenamefont {{Giblin}}, \citenamefont
  {{Kataoka}}, \citenamefont {{Fletcher}}, \citenamefont {{Chua}},
  \citenamefont {{Griffiths}}, \citenamefont {{Jones}}, \citenamefont
  {{Fal'ko}}, \citenamefont {{Smith}},\ and\ \citenamefont
  {{Janssen}}}]{npl_ghz_2013}%
  \BibitemOpen
  \bibfield  {author} {\bibinfo {author} {\bibfnamefont {M.~R.}\ \bibnamefont
  {{Connolly}}}, \bibinfo {author} {\bibfnamefont {K.~L.}\ \bibnamefont
  {{Chiu}}}, \bibinfo {author} {\bibfnamefont {S.~P.}\ \bibnamefont
  {{Giblin}}}, \bibinfo {author} {\bibfnamefont {M.}~\bibnamefont {{Kataoka}}},
  \bibinfo {author} {\bibfnamefont {J.~D.}\ \bibnamefont {{Fletcher}}},
  \bibinfo {author} {\bibfnamefont {C.}~\bibnamefont {{Chua}}}, \bibinfo
  {author} {\bibfnamefont {J.~P.}\ \bibnamefont {{Griffiths}}}, \bibinfo
  {author} {\bibfnamefont {G.~A.~C.}\ \bibnamefont {{Jones}}}, \bibinfo
  {author} {\bibfnamefont {V.~I.}\ \bibnamefont {{Fal'ko}}}, \bibinfo {author}
  {\bibfnamefont {C.~G.}\ \bibnamefont {{Smith}}}, \ and\ \bibinfo {author}
  {\bibfnamefont {T.~J.~B.~M.}\ \bibnamefont {{Janssen}}},\ }\href {\doibase
  10.1038/nnano.2013.73} {\bibfield  {journal} {\bibinfo  {journal} {Nature
  Nanotechnology}\ }\textbf {\bibinfo {volume} {8}},\ \bibinfo {pages} {417}
  (\bibinfo {year} {2013})}\BibitemShut {NoStop}%
\bibitem [{\citenamefont {Dubois}\ \emph {et~al.}(2013)\citenamefont {Dubois},
  \citenamefont {Jullien}, \citenamefont {Grenier}, \citenamefont {Degiovanni},
  \citenamefont {Roulleau},\ and\ \citenamefont {Glattli}}]{dubois_2013}%
  \BibitemOpen
  \bibfield  {author} {\bibinfo {author} {\bibfnamefont {J.}~\bibnamefont
  {Dubois}}, \bibinfo {author} {\bibfnamefont {T.}~\bibnamefont {Jullien}},
  \bibinfo {author} {\bibfnamefont {C.}~\bibnamefont {Grenier}}, \bibinfo
  {author} {\bibfnamefont {P.}~\bibnamefont {Degiovanni}}, \bibinfo {author}
  {\bibfnamefont {P.}~\bibnamefont {Roulleau}}, \ and\ \bibinfo {author}
  {\bibfnamefont {D.~C.}\ \bibnamefont {Glattli}},\ }\href {\doibase
  10.1103/PhysRevB.88.085301} {\bibfield  {journal} {\bibinfo  {journal} {Phys.
  Rev. B}\ }\textbf {\bibinfo {volume} {88}},\ \bibinfo {pages} {085301}
  (\bibinfo {year} {2013})}\BibitemShut {NoStop}%
\bibitem [{\citenamefont {{Ubbelohde}}\ \emph {et~al.}(2015)\citenamefont
  {{Ubbelohde}}, \citenamefont {{Hohls}}, \citenamefont {{Kashcheyevs}},
  \citenamefont {{Wagner}}, \citenamefont {{Fricke}}, \citenamefont
  {{K{\"a}stner}}, \citenamefont {{Pierz}}, \citenamefont {{Schumacher}},\ and\
  \citenamefont {{Haug}}}]{ubbelohde_2015}%
  \BibitemOpen
  \bibfield  {author} {\bibinfo {author} {\bibfnamefont {N.}~\bibnamefont
  {{Ubbelohde}}}, \bibinfo {author} {\bibfnamefont {F.}~\bibnamefont
  {{Hohls}}}, \bibinfo {author} {\bibfnamefont {V.}~\bibnamefont
  {{Kashcheyevs}}}, \bibinfo {author} {\bibfnamefont {T.}~\bibnamefont
  {{Wagner}}}, \bibinfo {author} {\bibfnamefont {L.}~\bibnamefont {{Fricke}}},
  \bibinfo {author} {\bibfnamefont {B.}~\bibnamefont {{K{\"a}stner}}}, \bibinfo
  {author} {\bibfnamefont {K.}~\bibnamefont {{Pierz}}}, \bibinfo {author}
  {\bibfnamefont {H.~W.}\ \bibnamefont {{Schumacher}}}, \ and\ \bibinfo
  {author} {\bibfnamefont {R.~J.}\ \bibnamefont {{Haug}}},\ }\href {\doibase
  10.1038/nnano.2014.275} {\bibfield  {journal} {\bibinfo  {journal} {Nature
  Nanotechnology}\ }\textbf {\bibinfo {volume} {10}},\ \bibinfo {pages} {46}
  (\bibinfo {year} {2015})}\BibitemShut {NoStop}%
\bibitem [{\citenamefont {{d'Hollosy}}\ \emph {et~al.}(2015)\citenamefont
  {{d'Hollosy}}, \citenamefont {{Jung}}, \citenamefont {{Baumgartner}},
  \citenamefont {{Guzenko}}, \citenamefont {{Madsen}}, \citenamefont
  {{Nyg{\aa}rd}},\ and\ \citenamefont {{Sch{\"o}nenberger}}}]{dHollsoy_2015}%
  \BibitemOpen
  \bibfield  {author} {\bibinfo {author} {\bibfnamefont {S.}~\bibnamefont
  {{d'Hollosy}}}, \bibinfo {author} {\bibfnamefont {M.}~\bibnamefont {{Jung}}},
  \bibinfo {author} {\bibfnamefont {A.}~\bibnamefont {{Baumgartner}}}, \bibinfo
  {author} {\bibfnamefont {V.~A.}\ \bibnamefont {{Guzenko}}}, \bibinfo {author}
  {\bibfnamefont {M.~H.}\ \bibnamefont {{Madsen}}}, \bibinfo {author}
  {\bibfnamefont {J.}~\bibnamefont {{Nyg{\aa}rd}}}, \ and\ \bibinfo {author}
  {\bibfnamefont {C.}~\bibnamefont {{Sch{\"o}nenberger}}},\ }\href {\doibase
  10.1021/acs.nanolett.5b01190} {\bibfield  {journal} {\bibinfo  {journal}
  {Nano Letters}\ }\textbf {\bibinfo {volume} {15}},\ \bibinfo {pages} {4585}
  (\bibinfo {year} {2015})}\BibitemShut {NoStop}%
\bibitem [{\citenamefont {Roussely}\ \emph {et~al.}(2018)\citenamefont
  {Roussely}, \citenamefont {Arrighi}, \citenamefont {Georgiou}, \citenamefont
  {Takada}, \citenamefont {Schalk}, \citenamefont {Urdampilleta}, \citenamefont
  {Ludwig}, \citenamefont {Wieck}, \citenamefont {Armagnat}, \citenamefont
  {Kloss}, \citenamefont {Waintal}, \citenamefont {Meunier},\ and\
  \citenamefont {B{\"{a}}uerle}}]{Roussely2018}%
  \BibitemOpen
  \bibfield  {author} {\bibinfo {author} {\bibfnamefont {G.}~\bibnamefont
  {Roussely}}, \bibinfo {author} {\bibfnamefont {E.}~\bibnamefont {Arrighi}},
  \bibinfo {author} {\bibfnamefont {G.}~\bibnamefont {Georgiou}}, \bibinfo
  {author} {\bibfnamefont {S.}~\bibnamefont {Takada}}, \bibinfo {author}
  {\bibfnamefont {M.}~\bibnamefont {Schalk}}, \bibinfo {author} {\bibfnamefont
  {M.}~\bibnamefont {Urdampilleta}}, \bibinfo {author} {\bibfnamefont
  {A.}~\bibnamefont {Ludwig}}, \bibinfo {author} {\bibfnamefont {A.~D.}\
  \bibnamefont {Wieck}}, \bibinfo {author} {\bibfnamefont {P.}~\bibnamefont
  {Armagnat}}, \bibinfo {author} {\bibfnamefont {T.}~\bibnamefont {Kloss}},
  \bibinfo {author} {\bibfnamefont {X.}~\bibnamefont {Waintal}}, \bibinfo
  {author} {\bibfnamefont {T.}~\bibnamefont {Meunier}}, \ and\ \bibinfo
  {author} {\bibfnamefont {C.}~\bibnamefont {B{\"{a}}uerle}},\ }\href {\doibase
  10.1038/s41467-018-05203-7} {\bibfield  {journal} {\bibinfo  {journal} {Nat.
  Commun.}\ }\textbf {\bibinfo {volume} {9}},\ \bibinfo {pages} {2811}
  (\bibinfo {year} {2018})}\BibitemShut {NoStop}%
\bibitem [{\citenamefont {Henny}\ \emph {et~al.}(1999)\citenamefont {Henny},
  \citenamefont {Oberholzer}, \citenamefont {Strunk}, \citenamefont {Heinzel},
  \citenamefont {Ensslin}, \citenamefont {Holland},\ and\ \citenamefont
  {Sch{\"o}nenberger}}]{henny_1999}%
  \BibitemOpen
  \bibfield  {author} {\bibinfo {author} {\bibfnamefont {M.}~\bibnamefont
  {Henny}}, \bibinfo {author} {\bibfnamefont {S.}~\bibnamefont {Oberholzer}},
  \bibinfo {author} {\bibfnamefont {C.}~\bibnamefont {Strunk}}, \bibinfo
  {author} {\bibfnamefont {T.}~\bibnamefont {Heinzel}}, \bibinfo {author}
  {\bibfnamefont {K.}~\bibnamefont {Ensslin}}, \bibinfo {author} {\bibfnamefont
  {M.}~\bibnamefont {Holland}}, \ and\ \bibinfo {author} {\bibfnamefont
  {C.}~\bibnamefont {Sch{\"o}nenberger}},\ }\href {\doibase
  10.1126/science.284.5412.296} {\bibfield  {journal} {\bibinfo  {journal}
  {Science}\ }\textbf {\bibinfo {volume} {284}},\ \bibinfo {pages} {296}
  (\bibinfo {year} {1999})}\BibitemShut {NoStop}%
\bibitem [{\citenamefont {Oliver}\ \emph {et~al.}(1999)\citenamefont {Oliver},
  \citenamefont {Kim}, \citenamefont {Liu},\ and\ \citenamefont
  {Yamamoto}}]{oliver_1999}%
  \BibitemOpen
  \bibfield  {author} {\bibinfo {author} {\bibfnamefont {W.~D.}\ \bibnamefont
  {Oliver}}, \bibinfo {author} {\bibfnamefont {J.}~\bibnamefont {Kim}},
  \bibinfo {author} {\bibfnamefont {R.~C.}\ \bibnamefont {Liu}}, \ and\
  \bibinfo {author} {\bibfnamefont {Y.}~\bibnamefont {Yamamoto}},\ }\href
  {\doibase 10.1126/science.284.5412.299} {\bibfield  {journal} {\bibinfo
  {journal} {Science}\ }\textbf {\bibinfo {volume} {284}},\ \bibinfo {pages}
  {299} (\bibinfo {year} {1999})}\BibitemShut {NoStop}%
\bibitem [{\citenamefont {{Kiesel}}\ \emph {et~al.}(2002)\citenamefont
  {{Kiesel}}, \citenamefont {{Renz}},\ and\ \citenamefont
  {{Hasselbach}}}]{kiesel_2002}%
  \BibitemOpen
  \bibfield  {author} {\bibinfo {author} {\bibfnamefont {H.}~\bibnamefont
  {{Kiesel}}}, \bibinfo {author} {\bibfnamefont {A.}~\bibnamefont {{Renz}}}, \
  and\ \bibinfo {author} {\bibfnamefont {F.}~\bibnamefont {{Hasselbach}}},\
  }\href {\doibase 10.1038/nature00911} {\bibfield  {journal} {\bibinfo
  {journal} {Nature}\ }\textbf {\bibinfo {volume} {418}},\ \bibinfo {pages}
  {392} (\bibinfo {year} {2002})}\BibitemShut {NoStop}%
\bibitem [{\citenamefont {Bocquillon}\ \emph {et~al.}(2012)\citenamefont
  {Bocquillon}, \citenamefont {Parmentier}, \citenamefont {Grenier},
  \citenamefont {Berroir}, \citenamefont {Degiovanni}, \citenamefont {Glattli},
  \citenamefont {Pla\ifmmode~\mbox{\c{c}}\else \c{c}\fi{}ais}, \citenamefont
  {Cavanna}, \citenamefont {Jin},\ and\ \citenamefont
  {F\`eve}}]{bocquillon_2012}%
  \BibitemOpen
  \bibfield  {author} {\bibinfo {author} {\bibfnamefont {E.}~\bibnamefont
  {Bocquillon}}, \bibinfo {author} {\bibfnamefont {F.~D.}\ \bibnamefont
  {Parmentier}}, \bibinfo {author} {\bibfnamefont {C.}~\bibnamefont {Grenier}},
  \bibinfo {author} {\bibfnamefont {J.-M.}\ \bibnamefont {Berroir}}, \bibinfo
  {author} {\bibfnamefont {P.}~\bibnamefont {Degiovanni}}, \bibinfo {author}
  {\bibfnamefont {D.~C.}\ \bibnamefont {Glattli}}, \bibinfo {author}
  {\bibfnamefont {B.}~\bibnamefont {Pla\ifmmode~\mbox{\c{c}}\else
  \c{c}\fi{}ais}}, \bibinfo {author} {\bibfnamefont {A.}~\bibnamefont
  {Cavanna}}, \bibinfo {author} {\bibfnamefont {Y.}~\bibnamefont {Jin}}, \ and\
  \bibinfo {author} {\bibfnamefont {G.}~\bibnamefont {F\`eve}},\ }\href
  {\doibase 10.1103/PhysRevLett.108.196803} {\bibfield  {journal} {\bibinfo
  {journal} {Phys. Rev. Lett.}\ }\textbf {\bibinfo {volume} {108}},\ \bibinfo
  {pages} {196803} (\bibinfo {year} {2012})}\BibitemShut {NoStop}%
\bibitem [{\citenamefont {Bocquillon}\ \emph {et~al.}(2013)\citenamefont
  {Bocquillon}, \citenamefont {Freulon}, \citenamefont {Berroir}, \citenamefont
  {Degiovanni}, \citenamefont {Pla{\c{c}}ais}, \citenamefont {Cavanna},
  \citenamefont {Jin},\ and\ \citenamefont {F{\`{e}}ve}}]{Bocquillon2013}%
  \BibitemOpen
  \bibfield  {author} {\bibinfo {author} {\bibfnamefont {E.}~\bibnamefont
  {Bocquillon}}, \bibinfo {author} {\bibfnamefont {V.}~\bibnamefont {Freulon}},
  \bibinfo {author} {\bibfnamefont {J.-M.}\ \bibnamefont {Berroir}}, \bibinfo
  {author} {\bibfnamefont {P.}~\bibnamefont {Degiovanni}}, \bibinfo {author}
  {\bibfnamefont {B.}~\bibnamefont {Pla{\c{c}}ais}}, \bibinfo {author}
  {\bibfnamefont {A.}~\bibnamefont {Cavanna}}, \bibinfo {author} {\bibfnamefont
  {Y.}~\bibnamefont {Jin}}, \ and\ \bibinfo {author} {\bibfnamefont
  {G.}~\bibnamefont {F{\`{e}}ve}},\ }\href {\doibase 10.1126/science.1232572}
  {\bibfield  {journal} {\bibinfo  {journal} {Science}\ }\textbf {\bibinfo
  {volume} {339}},\ \bibinfo {pages} {1054} (\bibinfo {year}
  {2013})}\BibitemShut {NoStop}%
\bibitem [{\citenamefont {Ferraro}\ \emph {et~al.}(2018)\citenamefont
  {Ferraro}, \citenamefont {Ronetti}, \citenamefont {Vannucci}, \citenamefont
  {Acciai}, \citenamefont {Rech}, \citenamefont {Jockheere}, \citenamefont
  {Martin},\ and\ \citenamefont {Sassetti}}]{Ferraro2018}%
  \BibitemOpen
  \bibfield  {author} {\bibinfo {author} {\bibfnamefont {D.}~\bibnamefont
  {Ferraro}}, \bibinfo {author} {\bibfnamefont {F.}~\bibnamefont {Ronetti}},
  \bibinfo {author} {\bibfnamefont {L.}~\bibnamefont {Vannucci}}, \bibinfo
  {author} {\bibfnamefont {M.}~\bibnamefont {Acciai}}, \bibinfo {author}
  {\bibfnamefont {J.}~\bibnamefont {Rech}}, \bibinfo {author} {\bibfnamefont
  {T.}~\bibnamefont {Jockheere}}, \bibinfo {author} {\bibfnamefont
  {T.}~\bibnamefont {Martin}}, \ and\ \bibinfo {author} {\bibfnamefont
  {M.}~\bibnamefont {Sassetti}},\ }\href {\doibase
  10.1140/epjst/e2018-800074-1} {\bibfield  {journal} {\bibinfo  {journal}
  {Eur. Phys. J. Spec. Top.}\ }\textbf {\bibinfo {volume} {227}},\ \bibinfo
  {pages} {1345} (\bibinfo {year} {2018})}\BibitemShut {NoStop}%
\bibitem [{\citenamefont {{Ji}}\ \emph {et~al.}(2003)\citenamefont {{Ji}},
  \citenamefont {{Chung}}, \citenamefont {{Sprinzak}}, \citenamefont
  {{Heiblum}}, \citenamefont {{Mahalu}},\ and\ \citenamefont
  {{Shtrikman}}}]{heiblum_2003}%
  \BibitemOpen
  \bibfield  {author} {\bibinfo {author} {\bibfnamefont {Y.}~\bibnamefont
  {{Ji}}}, \bibinfo {author} {\bibfnamefont {Y.}~\bibnamefont {{Chung}}},
  \bibinfo {author} {\bibfnamefont {D.}~\bibnamefont {{Sprinzak}}}, \bibinfo
  {author} {\bibfnamefont {M.}~\bibnamefont {{Heiblum}}}, \bibinfo {author}
  {\bibfnamefont {D.}~\bibnamefont {{Mahalu}}}, \ and\ \bibinfo {author}
  {\bibfnamefont {H.}~\bibnamefont {{Shtrikman}}},\ }\href {\doibase
  10.1038/nature01503} {\bibfield  {journal} {\bibinfo  {journal} {Nature}\
  }\textbf {\bibinfo {volume} {422}},\ \bibinfo {pages} {415} (\bibinfo {year}
  {2003})}\BibitemShut {NoStop}%
\bibitem [{\citenamefont {Neder}\ \emph {et~al.}(2006)\citenamefont {Neder},
  \citenamefont {Heiblum}, \citenamefont {Levinson}, \citenamefont {Mahalu},\
  and\ \citenamefont {Umansky}}]{heiblum_2006}%
  \BibitemOpen
  \bibfield  {author} {\bibinfo {author} {\bibfnamefont {I.}~\bibnamefont
  {Neder}}, \bibinfo {author} {\bibfnamefont {M.}~\bibnamefont {Heiblum}},
  \bibinfo {author} {\bibfnamefont {Y.}~\bibnamefont {Levinson}}, \bibinfo
  {author} {\bibfnamefont {D.}~\bibnamefont {Mahalu}}, \ and\ \bibinfo {author}
  {\bibfnamefont {V.}~\bibnamefont {Umansky}},\ }\href {\doibase
  10.1103/PhysRevLett.96.016804} {\bibfield  {journal} {\bibinfo  {journal}
  {Phys. Rev. Lett.}\ }\textbf {\bibinfo {volume} {96}},\ \bibinfo {pages}
  {016804} (\bibinfo {year} {2006})}\BibitemShut {NoStop}%
\bibitem [{\citenamefont {Neder}\ \emph {et~al.}(2007)\citenamefont {Neder},
  \citenamefont {Heiblum}, \citenamefont {Mahalu},\ and\ \citenamefont
  {Umansky}}]{heiblum_2007}%
  \BibitemOpen
  \bibfield  {author} {\bibinfo {author} {\bibfnamefont {I.}~\bibnamefont
  {Neder}}, \bibinfo {author} {\bibfnamefont {M.}~\bibnamefont {Heiblum}},
  \bibinfo {author} {\bibfnamefont {D.}~\bibnamefont {Mahalu}}, \ and\ \bibinfo
  {author} {\bibfnamefont {V.}~\bibnamefont {Umansky}},\ }\href {\doibase
  10.1103/PhysRevLett.98.036803} {\bibfield  {journal} {\bibinfo  {journal}
  {Phys. Rev. Lett.}\ }\textbf {\bibinfo {volume} {98}},\ \bibinfo {pages}
  {036803} (\bibinfo {year} {2007})}\BibitemShut {NoStop}%
\bibitem [{\citenamefont {Bocquillon}\ \emph
  {et~al.}(2014{\natexlab{b}})\citenamefont {Bocquillon}, \citenamefont
  {Freulon}, \citenamefont {Parmentier}, \citenamefont {Berroir}, \citenamefont
  {Pla{\c{c}}ais}, \citenamefont {Wahl}, \citenamefont {Rech}, \citenamefont
  {Jonckheere}, \citenamefont {Martin}, \citenamefont {Grenier}, \citenamefont
  {Ferraro}, \citenamefont {Degiovanni},\ and\ \citenamefont
  {F{\`{e}}ve}}]{Bocquillon2014}%
  \BibitemOpen
  \bibfield  {author} {\bibinfo {author} {\bibfnamefont {E.}~\bibnamefont
  {Bocquillon}}, \bibinfo {author} {\bibfnamefont {V.}~\bibnamefont {Freulon}},
  \bibinfo {author} {\bibfnamefont {F.~D.}\ \bibnamefont {Parmentier}},
  \bibinfo {author} {\bibfnamefont {J.-M.}\ \bibnamefont {Berroir}}, \bibinfo
  {author} {\bibfnamefont {B.}~\bibnamefont {Pla{\c{c}}ais}}, \bibinfo {author}
  {\bibfnamefont {C.}~\bibnamefont {Wahl}}, \bibinfo {author} {\bibfnamefont
  {J.}~\bibnamefont {Rech}}, \bibinfo {author} {\bibfnamefont {T.}~\bibnamefont
  {Jonckheere}}, \bibinfo {author} {\bibfnamefont {T.}~\bibnamefont {Martin}},
  \bibinfo {author} {\bibfnamefont {C.}~\bibnamefont {Grenier}}, \bibinfo
  {author} {\bibfnamefont {D.}~\bibnamefont {Ferraro}}, \bibinfo {author}
  {\bibfnamefont {P.}~\bibnamefont {Degiovanni}}, \ and\ \bibinfo {author}
  {\bibfnamefont {G.}~\bibnamefont {F{\`{e}}ve}},\ }\href {\doibase
  10.1002/andp.201300181} {\bibfield  {journal} {\bibinfo  {journal} {Ann.
  Phys.}\ }\textbf {\bibinfo {volume} {526}},\ \bibinfo {pages} {1} (\bibinfo
  {year} {2014}{\natexlab{b}})}\BibitemShut {NoStop}%
\bibitem [{\citenamefont {Grenier}\ \emph
  {et~al.}(2011{\natexlab{b}})\citenamefont {Grenier}, \citenamefont
  {Herv{\'{e}}}, \citenamefont {Bocquillon}, \citenamefont {Parmentier},
  \citenamefont {Pla{\c{c}}ais}, \citenamefont {Berroir}, \citenamefont
  {F{\`{e}}ve},\ and\ \citenamefont {Degiovanni}}]{Grenier2011}%
  \BibitemOpen
  \bibfield  {author} {\bibinfo {author} {\bibfnamefont {C.}~\bibnamefont
  {Grenier}}, \bibinfo {author} {\bibfnamefont {R.}~\bibnamefont
  {Herv{\'{e}}}}, \bibinfo {author} {\bibfnamefont {E.}~\bibnamefont
  {Bocquillon}}, \bibinfo {author} {\bibfnamefont {F.~D.}\ \bibnamefont
  {Parmentier}}, \bibinfo {author} {\bibfnamefont {B.}~\bibnamefont
  {Pla{\c{c}}ais}}, \bibinfo {author} {\bibfnamefont {J.~M.}\ \bibnamefont
  {Berroir}}, \bibinfo {author} {\bibfnamefont {G.}~\bibnamefont {F{\`{e}}ve}},
  \ and\ \bibinfo {author} {\bibfnamefont {P.}~\bibnamefont {Degiovanni}},\
  }\href {\doibase 10.1088/1367-2630/13/9/093007} {\bibfield  {journal}
  {\bibinfo  {journal} {New J. Phys.}\ }\textbf {\bibinfo {volume} {13}},\
  \bibinfo {pages} {093007} (\bibinfo {year} {2011}{\natexlab{b}})}\BibitemShut
  {NoStop}%
\bibitem [{\citenamefont {{Jullien}}\ \emph {et~al.}(2014)\citenamefont
  {{Jullien}}, \citenamefont {{Roulleau}}, \citenamefont {{Roche}},
  \citenamefont {{Cavanna}}, \citenamefont {{Jin}},\ and\ \citenamefont
  {{Glattli}}}]{glattli_2014}%
  \BibitemOpen
  \bibfield  {author} {\bibinfo {author} {\bibfnamefont {T.}~\bibnamefont
  {{Jullien}}}, \bibinfo {author} {\bibfnamefont {P.}~\bibnamefont
  {{Roulleau}}}, \bibinfo {author} {\bibfnamefont {B.}~\bibnamefont {{Roche}}},
  \bibinfo {author} {\bibfnamefont {A.}~\bibnamefont {{Cavanna}}}, \bibinfo
  {author} {\bibfnamefont {Y.}~\bibnamefont {{Jin}}}, \ and\ \bibinfo {author}
  {\bibfnamefont {D.~C.}\ \bibnamefont {{Glattli}}},\ }\href {\doibase
  10.1038/nature13821} {\bibfield  {journal} {\bibinfo  {journal} {Nature}\
  }\textbf {\bibinfo {volume} {514}},\ \bibinfo {pages} {603} (\bibinfo {year}
  {2014})}\BibitemShut {NoStop}%
\bibitem [{\citenamefont {Marguerite}\ \emph {et~al.}()\citenamefont
  {Marguerite}, \citenamefont {Roussel}, \citenamefont {Bisognin},
  \citenamefont {Cabart}, \citenamefont {Kumar}, \citenamefont {Berroir},
  \citenamefont {Bocquillon}, \citenamefont {Pla{\c{c}}ais}, \citenamefont
  {Cavanna}, \citenamefont {Gennser}, \citenamefont {Jin}, \citenamefont
  {Degiovanni},\ and\ \citenamefont {F{\`{e}}ve}}]{Marguerite2017}%
  \BibitemOpen
  \bibfield  {author} {\bibinfo {author} {\bibfnamefont {A.}~\bibnamefont
  {Marguerite}}, \bibinfo {author} {\bibfnamefont {B.}~\bibnamefont {Roussel}},
  \bibinfo {author} {\bibfnamefont {R.}~\bibnamefont {Bisognin}}, \bibinfo
  {author} {\bibfnamefont {C.}~\bibnamefont {Cabart}}, \bibinfo {author}
  {\bibfnamefont {M.}~\bibnamefont {Kumar}}, \bibinfo {author} {\bibfnamefont
  {J.~M.}\ \bibnamefont {Berroir}}, \bibinfo {author} {\bibfnamefont
  {E.}~\bibnamefont {Bocquillon}}, \bibinfo {author} {\bibfnamefont
  {B.}~\bibnamefont {Pla{\c{c}}ais}}, \bibinfo {author} {\bibfnamefont
  {A.}~\bibnamefont {Cavanna}}, \bibinfo {author} {\bibfnamefont
  {U.}~\bibnamefont {Gennser}}, \bibinfo {author} {\bibfnamefont
  {Y.}~\bibnamefont {Jin}}, \bibinfo {author} {\bibfnamefont {P.}~\bibnamefont
  {Degiovanni}}, \ and\ \bibinfo {author} {\bibfnamefont {G.}~\bibnamefont
  {F{\`{e}}ve}},\ }\href {http://arxiv.org/abs/1710.11181} {\enquote {\bibinfo
  {title} {{Extracting single electron wavefunctions from a quantum electrical
  current}},}\ }\Eprint {http://arxiv.org/abs/1710.11181} {arXiv:1710.11181}
  \BibitemShut {NoStop}%
\bibitem [{\citenamefont {Leicht}\ \emph {et~al.}(2011)\citenamefont {Leicht},
  \citenamefont {Mirovsky}, \citenamefont {Kaestner}, \citenamefont {Hohls},
  \citenamefont {Kashcheyevs}, \citenamefont {Kurganova}, \citenamefont
  {Zeitler}, \citenamefont {Weimann}, \citenamefont {Pierz},\ and\
  \citenamefont {Schumacher}}]{kaestner2010d}%
  \BibitemOpen
  \bibfield  {author} {\bibinfo {author} {\bibfnamefont {C.}~\bibnamefont
  {Leicht}}, \bibinfo {author} {\bibfnamefont {P.}~\bibnamefont {Mirovsky}},
  \bibinfo {author} {\bibfnamefont {B.}~\bibnamefont {Kaestner}}, \bibinfo
  {author} {\bibfnamefont {F.}~\bibnamefont {Hohls}}, \bibinfo {author}
  {\bibfnamefont {V.}~\bibnamefont {Kashcheyevs}}, \bibinfo {author}
  {\bibfnamefont {E.~V.}\ \bibnamefont {Kurganova}}, \bibinfo {author}
  {\bibfnamefont {U.}~\bibnamefont {Zeitler}}, \bibinfo {author} {\bibfnamefont
  {T.}~\bibnamefont {Weimann}}, \bibinfo {author} {\bibfnamefont
  {K.}~\bibnamefont {Pierz}}, \ and\ \bibinfo {author} {\bibfnamefont {H.~W.}\
  \bibnamefont {Schumacher}},\ }\href {\doibase 10.1088/0268-1242/26/5/055010}
  {\bibfield  {journal} {\bibinfo  {journal} {Semicond. Sci. Technol.}\
  }\textbf {\bibinfo {volume} {26}},\ \bibinfo {pages} {055010} (\bibinfo
  {year} {2011})}\BibitemShut {NoStop}%
\bibitem [{\citenamefont {Fletcher}\ \emph {et~al.}(2013)\citenamefont
  {Fletcher}, \citenamefont {See}, \citenamefont {Howe}, \citenamefont
  {Pepper}, \citenamefont {Giblin}, \citenamefont {Griffiths}, \citenamefont
  {Jones}, \citenamefont {Farrer}, \citenamefont {Ritchie}, \citenamefont
  {Janssen},\ and\ \citenamefont {Kataoka}}]{fletcher_detect_2013}%
  \BibitemOpen
  \bibfield  {author} {\bibinfo {author} {\bibfnamefont {J.~D.}\ \bibnamefont
  {Fletcher}}, \bibinfo {author} {\bibfnamefont {P.}~\bibnamefont {See}},
  \bibinfo {author} {\bibfnamefont {H.}~\bibnamefont {Howe}}, \bibinfo {author}
  {\bibfnamefont {M.}~\bibnamefont {Pepper}}, \bibinfo {author} {\bibfnamefont
  {S.~P.}\ \bibnamefont {Giblin}}, \bibinfo {author} {\bibfnamefont {J.~P.}\
  \bibnamefont {Griffiths}}, \bibinfo {author} {\bibfnamefont {G.~A.~C.}\
  \bibnamefont {Jones}}, \bibinfo {author} {\bibfnamefont {I.}~\bibnamefont
  {Farrer}}, \bibinfo {author} {\bibfnamefont {D.~A.}\ \bibnamefont {Ritchie}},
  \bibinfo {author} {\bibfnamefont {T.~J. B.~M.}\ \bibnamefont {Janssen}}, \
  and\ \bibinfo {author} {\bibfnamefont {M.}~\bibnamefont {Kataoka}},\ }\href
  {\doibase 10.1103/PhysRevLett.111.216807} {\bibfield  {journal} {\bibinfo
  {journal} {Phys. Rev. Lett.}\ }\textbf {\bibinfo {volume} {111}},\ \bibinfo
  {pages} {216807} (\bibinfo {year} {2013})}\BibitemShut {NoStop}%
\bibitem [{\citenamefont {Waldie}\ \emph {et~al.}(2015)\citenamefont {Waldie},
  \citenamefont {See}, \citenamefont {Kashcheyevs}, \citenamefont {Griffiths},
  \citenamefont {Farrer}, \citenamefont {Jones}, \citenamefont {Ritchie},
  \citenamefont {Janssen},\ and\ \citenamefont {Kataoka}}]{waldie_detect_2015}%
  \BibitemOpen
  \bibfield  {author} {\bibinfo {author} {\bibfnamefont {J.}~\bibnamefont
  {Waldie}}, \bibinfo {author} {\bibfnamefont {P.}~\bibnamefont {See}},
  \bibinfo {author} {\bibfnamefont {V.}~\bibnamefont {Kashcheyevs}}, \bibinfo
  {author} {\bibfnamefont {J.~P.}\ \bibnamefont {Griffiths}}, \bibinfo {author}
  {\bibfnamefont {I.}~\bibnamefont {Farrer}}, \bibinfo {author} {\bibfnamefont
  {G.~A.~C.}\ \bibnamefont {Jones}}, \bibinfo {author} {\bibfnamefont {D.~A.}\
  \bibnamefont {Ritchie}}, \bibinfo {author} {\bibfnamefont {T.~J. B.~M.}\
  \bibnamefont {Janssen}}, \ and\ \bibinfo {author} {\bibfnamefont
  {M.}~\bibnamefont {Kataoka}},\ }\href {\doibase 10.1103/PhysRevB.92.125305}
  {\bibfield  {journal} {\bibinfo  {journal} {Phys. Rev. B}\ }\textbf {\bibinfo
  {volume} {92}},\ \bibinfo {pages} {125305} (\bibinfo {year}
  {2015})}\BibitemShut {NoStop}%
\bibitem [{\citenamefont {Kataoka}\ \emph {et~al.}(2016)\citenamefont
  {Kataoka}, \citenamefont {Johnson}, \citenamefont {Emary}, \citenamefont
  {See}, \citenamefont {Griffiths}, \citenamefont {Jones}, \citenamefont
  {Farrer}, \citenamefont {Ritchie}, \citenamefont {Pepper},\ and\
  \citenamefont {Janssen}}]{kataoka_detect_2016}%
  \BibitemOpen
  \bibfield  {author} {\bibinfo {author} {\bibfnamefont {M.}~\bibnamefont
  {Kataoka}}, \bibinfo {author} {\bibfnamefont {N.}~\bibnamefont {Johnson}},
  \bibinfo {author} {\bibfnamefont {C.}~\bibnamefont {Emary}}, \bibinfo
  {author} {\bibfnamefont {P.}~\bibnamefont {See}}, \bibinfo {author}
  {\bibfnamefont {J.~P.}\ \bibnamefont {Griffiths}}, \bibinfo {author}
  {\bibfnamefont {G.~A.~C.}\ \bibnamefont {Jones}}, \bibinfo {author}
  {\bibfnamefont {I.}~\bibnamefont {Farrer}}, \bibinfo {author} {\bibfnamefont
  {D.~A.}\ \bibnamefont {Ritchie}}, \bibinfo {author} {\bibfnamefont
  {M.}~\bibnamefont {Pepper}}, \ and\ \bibinfo {author} {\bibfnamefont {T.~J.
  B.~M.}\ \bibnamefont {Janssen}},\ }\href {\doibase
  10.1103/PhysRevLett.116.126803} {\bibfield  {journal} {\bibinfo  {journal}
  {Phys. Rev. Lett.}\ }\textbf {\bibinfo {volume} {116}},\ \bibinfo {pages}
  {126803} (\bibinfo {year} {2016})}\BibitemShut {NoStop}%
\bibitem [{\citenamefont {Kataoka}\ \emph {et~al.}(2017)\citenamefont
  {Kataoka}, \citenamefont {Fletcher},\ and\ \citenamefont
  {Johnson}}]{kataoka_pss}%
  \BibitemOpen
  \bibfield  {author} {\bibinfo {author} {\bibfnamefont {M.}~\bibnamefont
  {Kataoka}}, \bibinfo {author} {\bibfnamefont {J.~D.}\ \bibnamefont
  {Fletcher}}, \ and\ \bibinfo {author} {\bibfnamefont {N.}~\bibnamefont
  {Johnson}},\ }\href {\doibase 10.1002/pssb.201600547} {\bibfield  {journal}
  {\bibinfo  {journal} {Phys. Status Solidi B}\ }\textbf {\bibinfo {volume}
  {254}},\ \bibinfo {pages} {1600547} (\bibinfo {year} {2017})}\BibitemShut
  {NoStop}%
\bibitem [{\citenamefont {Fletcher}\ \emph {et~al.}()\citenamefont {Fletcher},
  \citenamefont {Johnson}, \citenamefont {Locane}, \citenamefont {See},
  \citenamefont {Griffiths}, \citenamefont {Farrer}, \citenamefont {Ritchie},
  \citenamefont {Brouwer}, \citenamefont {Kashcheyevs},\ and\ \citenamefont
  {Kataoka}}]{fletcher_2018}%
  \BibitemOpen
  \bibfield  {author} {\bibinfo {author} {\bibfnamefont {J.~D.}\ \bibnamefont
  {Fletcher}}, \bibinfo {author} {\bibfnamefont {N.}~\bibnamefont {Johnson}},
  \bibinfo {author} {\bibfnamefont {E.}~\bibnamefont {Locane}}, \bibinfo
  {author} {\bibfnamefont {P.}~\bibnamefont {See}}, \bibinfo {author}
  {\bibfnamefont {J.~P.}\ \bibnamefont {Griffiths}}, \bibinfo {author}
  {\bibfnamefont {I.}~\bibnamefont {Farrer}}, \bibinfo {author} {\bibfnamefont
  {D.~A.}\ \bibnamefont {Ritchie}}, \bibinfo {author} {\bibfnamefont {P.~W.}\
  \bibnamefont {Brouwer}}, \bibinfo {author} {\bibfnamefont {V.}~\bibnamefont
  {Kashcheyevs}}, \ and\ \bibinfo {author} {\bibfnamefont {M.}~\bibnamefont
  {Kataoka}},\ }\href {http://arxiv.org/abs/1901.10985} {\enquote {\bibinfo
  {title} {{Quantum Tomography of Solitary Electrons}},}\ }\Eprint
  {http://arxiv.org/abs/1901.10985} {arXiv:1901.10985} \BibitemShut {NoStop}%
\bibitem [{\citenamefont {Ferraro}\ \emph {et~al.}(2013)\citenamefont
  {Ferraro}, \citenamefont {Feller}, \citenamefont {Ghibaudo}, \citenamefont
  {Thibierge}, \citenamefont {Bocquillon}, \citenamefont {F\`eve},
  \citenamefont {Grenier},\ and\ \citenamefont {Degiovanni}}]{ferraro_2013}%
  \BibitemOpen
  \bibfield  {author} {\bibinfo {author} {\bibfnamefont {D.}~\bibnamefont
  {Ferraro}}, \bibinfo {author} {\bibfnamefont {A.}~\bibnamefont {Feller}},
  \bibinfo {author} {\bibfnamefont {A.}~\bibnamefont {Ghibaudo}}, \bibinfo
  {author} {\bibfnamefont {E.}~\bibnamefont {Thibierge}}, \bibinfo {author}
  {\bibfnamefont {E.}~\bibnamefont {Bocquillon}}, \bibinfo {author}
  {\bibfnamefont {G.}~\bibnamefont {F\`eve}}, \bibinfo {author} {\bibfnamefont
  {C.}~\bibnamefont {Grenier}}, \ and\ \bibinfo {author} {\bibfnamefont
  {P.}~\bibnamefont {Degiovanni}},\ }\href {\doibase
  10.1103/PhysRevB.88.205303} {\bibfield  {journal} {\bibinfo  {journal} {Phys.
  Rev. B}\ }\textbf {\bibinfo {volume} {88}},\ \bibinfo {pages} {205303}
  (\bibinfo {year} {2013})}\BibitemShut {NoStop}%
\bibitem [{\citenamefont {{Kashcheyevs}}\ and\ \citenamefont
  {{Samuelsson}}(2017)}]{slava_peter}%
  \BibitemOpen
  \bibfield  {author} {\bibinfo {author} {\bibfnamefont {V.}~\bibnamefont
  {{Kashcheyevs}}}\ and\ \bibinfo {author} {\bibfnamefont {P.}~\bibnamefont
  {{Samuelsson}}},\ }\href {\doibase 10.1103/PhysRevB.95.245424} {\bibfield
  {journal} {\bibinfo  {journal} {\prb}\ }\textbf {\bibinfo {volume} {95}},\
  \bibinfo {eid} {245424} (\bibinfo {year} {2017})}\BibitemShut {NoStop}%
\bibitem [{\citenamefont {Emary}\ \emph {et~al.}(2019)\citenamefont {Emary},
  \citenamefont {Clark}, \citenamefont {Kataoka},\ and\ \citenamefont
  {Johnson}}]{Emary2019}%
  \BibitemOpen
  \bibfield  {author} {\bibinfo {author} {\bibfnamefont {C.}~\bibnamefont
  {Emary}}, \bibinfo {author} {\bibfnamefont {L.~A.}\ \bibnamefont {Clark}},
  \bibinfo {author} {\bibfnamefont {M.}~\bibnamefont {Kataoka}}, \ and\
  \bibinfo {author} {\bibfnamefont {N.}~\bibnamefont {Johnson}},\ }\href
  {\doibase 10.1103/PhysRevB.99.045306} {\bibfield  {journal} {\bibinfo
  {journal} {Phys. Rev. B}\ }\textbf {\bibinfo {volume} {99}},\ \bibinfo
  {pages} {045306} (\bibinfo {year} {2019})}\BibitemShut {NoStop}%
\bibitem [{\citenamefont {Feit}\ \emph {et~al.}(1982)\citenamefont {Feit},
  \citenamefont {Fleck},\ and\ \citenamefont {Steiger}}]{Feit1982}%
  \BibitemOpen
  \bibfield  {author} {\bibinfo {author} {\bibfnamefont {M.}~\bibnamefont
  {Feit}}, \bibinfo {author} {\bibfnamefont {J.}~\bibnamefont {Fleck}}, \ and\
  \bibinfo {author} {\bibfnamefont {A.}~\bibnamefont {Steiger}},\ }\href
  {\doibase 10.1016/0021-9991(82)90091-2} {\bibfield  {journal} {\bibinfo
  {journal} {J. Comput. Phys.}\ }\textbf {\bibinfo {volume} {47}},\ \bibinfo
  {pages} {412} (\bibinfo {year} {1982})}\BibitemShut {NoStop}%
\bibitem [{\citenamefont {Bellentani}\ \emph {et~al.}(2018)\citenamefont
  {Bellentani}, \citenamefont {Beggi}, \citenamefont {Bordone},\ and\
  \citenamefont {Bertoni}}]{Bellentani2018}%
  \BibitemOpen
  \bibfield  {author} {\bibinfo {author} {\bibfnamefont {L.}~\bibnamefont
  {Bellentani}}, \bibinfo {author} {\bibfnamefont {A.}~\bibnamefont {Beggi}},
  \bibinfo {author} {\bibfnamefont {P.}~\bibnamefont {Bordone}}, \ and\
  \bibinfo {author} {\bibfnamefont {A.}~\bibnamefont {Bertoni}},\ }\href
  {\doibase 10.1103/PhysRevB.97.205419} {\bibfield  {journal} {\bibinfo
  {journal} {Phys. Rev. B}\ }\textbf {\bibinfo {volume} {97}},\ \bibinfo
  {pages} {205419} (\bibinfo {year} {2018})}\BibitemShut {NoStop}%
\bibitem [{\citenamefont {Aharony}\ and\ \citenamefont
  {Entin-Wohlman}(2002)}]{Aharony02PRL}%
  \BibitemOpen
  \bibfield  {author} {\bibinfo {author} {\bibfnamefont {A.}~\bibnamefont
  {Aharony}}\ and\ \bibinfo {author} {\bibfnamefont {O.}~\bibnamefont
  {Entin-Wohlman}},\ }\href {\doibase 10.1103/PhysRevB.65.241401} {\bibfield
  {journal} {\bibinfo  {journal} {Phys. Rev. B}\ }\textbf {\bibinfo {volume}
  {65}},\ \bibinfo {pages} {241401} (\bibinfo {year} {2002})}\BibitemShut
  {NoStop}%
\bibitem [{\citenamefont {Deans}(1992)}]{radon}%
  \BibitemOpen
  \bibfield  {author} {\bibinfo {author} {\bibfnamefont {S.~R.}\ \bibnamefont
  {Deans}},\ }\href@noop {} {\emph {\bibinfo {title} {The Radon Transform and
  Some of Its Applications}}}\ (\bibinfo  {publisher} {Krieger Pub Co},\
  \bibinfo {year} {1992})\BibitemShut {NoStop}%
\bibitem [{\citenamefont {Johnson}\ \emph {et~al.}(2017)\citenamefont
  {Johnson}, \citenamefont {Fletcher}, \citenamefont {Humphreys}, \citenamefont
  {See}, \citenamefont {Griffiths}, \citenamefont {Jones}, \citenamefont
  {Farrer}, \citenamefont {Ritchie}, \citenamefont {Pepper}, \citenamefont
  {Janssen},\ and\ \citenamefont {Kataoka}}]{johnson_2017}%
  \BibitemOpen
  \bibfield  {author} {\bibinfo {author} {\bibfnamefont {N.}~\bibnamefont
  {Johnson}}, \bibinfo {author} {\bibfnamefont {J.~D.}\ \bibnamefont
  {Fletcher}}, \bibinfo {author} {\bibfnamefont {D.~A.}\ \bibnamefont
  {Humphreys}}, \bibinfo {author} {\bibfnamefont {P.}~\bibnamefont {See}},
  \bibinfo {author} {\bibfnamefont {J.~P.}\ \bibnamefont {Griffiths}}, \bibinfo
  {author} {\bibfnamefont {G.~A.~C.}\ \bibnamefont {Jones}}, \bibinfo {author}
  {\bibfnamefont {I.}~\bibnamefont {Farrer}}, \bibinfo {author} {\bibfnamefont
  {D.~A.}\ \bibnamefont {Ritchie}}, \bibinfo {author} {\bibfnamefont
  {M.}~\bibnamefont {Pepper}}, \bibinfo {author} {\bibfnamefont {T.~J. B.~M.}\
  \bibnamefont {Janssen}}, \ and\ \bibinfo {author} {\bibfnamefont
  {M.}~\bibnamefont {Kataoka}},\ }\href {\doibase 10.1063/1.4978388} {\bibfield
   {journal} {\bibinfo  {journal} {Applied Physics Letters}\ }\textbf {\bibinfo
  {volume} {110}},\ \bibinfo {pages} {102105} (\bibinfo {year}
  {2017})}\BibitemShut {NoStop}%
\end{thebibliography}

%

\end{document}